\documentclass[
prb,twocolumn,superscriptaddress]{revtex4-2}
\usepackage{graphicx}
\usepackage{color,bm}
\usepackage{nameref}
\usepackage[colorlinks,bookmarks=false,citecolor=blue, urlcolor=blue]{hyperref}
\usepackage{amsmath,amssymb}
\usepackage{mathrsfs}
\usepackage{siunitx}
\usepackage[export]{adjustbox} 
\usepackage{multirow}
\usepackage{svg}
\usepackage{array} 
\usepackage{colortbl}

\begin{document}

\title{3D Magnetic Textures with Mixed Topology: Unlocking the Tunable Hopf Index}

\author{Maria Azhar}
\affiliation{Faculty of Physics and Center for Nanointegration Duisburg-Essen (CENIDE), University of Duisburg-Essen, 47057 Duisburg, Germany}

\author{Sandra C.\ Shaju}
\affiliation{Faculty of Physics and Center for Nanointegration Duisburg-Essen (CENIDE), University of Duisburg-Essen, 47057 Duisburg, Germany}

\author{Ross Knapman}
\affiliation{Faculty of Physics and Center for Nanointegration Duisburg-Essen (CENIDE), University of Duisburg-Essen, 47057 Duisburg, Germany}

\author{Alessandro Pignedoli}
\affiliation{Faculty of Physics and Center for Nanointegration Duisburg-Essen (CENIDE), University of Duisburg-Essen, 47057 Duisburg, Germany}

\author{Karin Everschor-Sitte}
\affiliation{Faculty of Physics and Center for Nanointegration Duisburg-Essen (CENIDE), University of Duisburg-Essen, 47057 Duisburg, Germany}

\date{\today}

\begin{abstract}
	Knots and links play a crucial role in understanding topology and discreteness in nature. In magnetic systems, twisted, knotted and braided vortex tubes manifest as Skyrmions, Hopfions, or screw dislocations. These complex textures are characterized by topologically non-trivial quantities, such as a Skyrmion number, a Hopf index $H$, a Burgers vector (quantified by an integer $\nu$), and linking numbers. In this work, we introduce a discrete geometric definition of $H$ for periodic magnetic textures, which can be separated into contributions from the self-linking and inter-linking of flux tubes. We show that fractional Hopfions or textures with non-integer values of $H$ naturally arise and can be interpreted as states of ``mixed topology" that are continuously transformable to one of the multiple possible topological sectors. Our findings demonstrate a solid physical foundation for the Hopf index to take integer, non-integer, or specific fractional values, depending on the underlying topology of the system.
\end{abstract}

\maketitle

\section{Introduction}
\begin{figure}[b!]
\includegraphics[width=1\columnwidth]{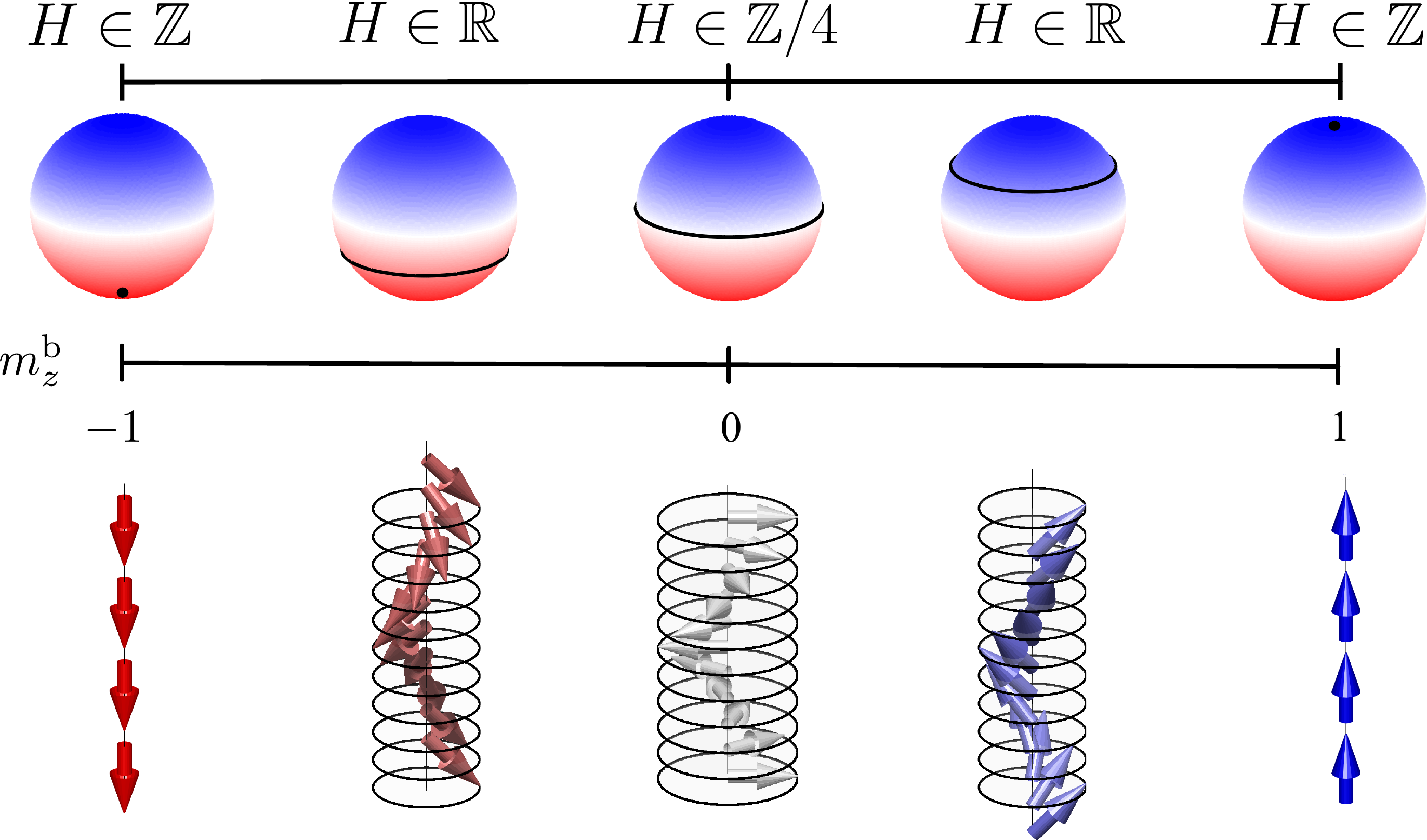}
	\caption{The Hopf index, $H$, can take fractional values when the background maps to an $S^1$ line within the target space $S^2$ of the magnetisation vector $\mathbf{m}$ (upper panel). The magnetisation of the background, $m_z^{\mathrm{b}}$ (depicted in the lower panel), determines the latitude of this line, effectively partitioning the $S^2$ sphere into two topologically distinct subspaces.}
	\label{fig:fig1}
\end{figure}

Knots, links, and braids are important across a wide range of scientific disciplines and are realisable in diverse physical platforms \cite{KnottedFields2024} such as water \cite{Kleckner2013,Neophytou2022}, quantum gases \cite{Ollikainen2019}, electromagnetic waves \cite{Kedia2013}, DNA \cite{Han2010} superfluids \cite{Kleckner2016}, liquid crystals \cite{Tkalec2011, Binysh2020,Smalyukh_2020, Wu2022}, laser light \cite{Sugic2021, Kong2022}, high energy physics \cite{Faddeev1997}, frustrated magnets \cite{Sutcliffe2017, Rybakov2022} and chiral magnets \cite{Kent2021,Tai18, Sutcliffe18, Voinescu2020, Azhar2022, Zheng2023}. These phenomena are significant across length scales, from the smallest lengths in the early cosmology of the universe \cite{Duan2004} to magnetic braids in the solar corona that store vast amounts of energy, making them several times hotter than the surface of the sun \cite{Cirtain2013}. 

In magnetic materials, a multitude of topologically non-trivial textures has been realised \cite{Gobel2021}, such as Skyrmions in 2D systems that can be readily manipulated with external fields and have potential in device applications \cite{Back_2020,  Fert2024, Liu2024, delSer2023}.  In 3D systems, the addition of a third spatial dimension enriches the applications of topology by enabling knots, links and braids, to form textures~\cite{Tang2021,Zhang2024} such as Hopfions~\cite{Sutcliffe2017, Barts2021} or the recently predicted \cite{Azhar2022} and observed \cite{Zheng2023} screw dislocations. 
Also, 3D textures exhibit complex dynamics under applied drives \cite{Wang2019, Gobel2020, Liu2020, Raftrey2021} making them interesting for applications.
Notably, the family of topological defects has been expanded to include textures where non-integer Hopf indices have been calculated~\cite{Yu2023, Zhang2024}. 

Magnetic textures have typically been characterised based on the homotopy groups of spheres or mapping of the magnetisation $\mathbf{m}$ from a base space to the target space $S^2$. The standard homotopy groups resulting in integer values of the Skyrmion number $N_{\mathrm{sk}}$ and Hopf index $H$ are $\pi_2(S^2)=\mathbb{Z}$ and $\pi_3(S^2)=\mathbb{Z}$ for 2D and 3D textures, respectively.
Topological stability is then derived from the fact that knotted textures in a continuous field cannot be unwound without introducing singularities, violating the continuity of the field or modifying the base space.

In this work, we show that a classification of 3D magnetic textures by homotopy groups is not sufficient to explain all classes of 3D topological magnetic defects such as Hopfions with non-integer Hopf indices. 
Instead, it is required to introduce the concept of flux tubes of the emergent magnetic field corresponding to the magnetic textures and analyse their linking numbers~\cite{Milnor1954, Moffatt2013, Machon2019, Copar2011, Scheeler2017} as topological invariants as well.
Using this concept, we demonstrate that arbitrary non-integer values of the Hopf index naturally arise for magnetic structures whose far-field or background magnetisation is non-collinear and maps to a $S^1$ line on the $S^2$ magnetisation sphere (see Fig.~\ref{fig:fig1}), despite a discrete representation of the Hopf index by a finite number of flux tubes. Examples of this are magnetic textures in a conical spiral~\cite{Leonov_2016, Rybakov2015, Voinescu2020, Kuchkin2023} or screw dislocation~\cite{Azhar2022} background.

Here we focus on magnetisation textures which are periodic along a certain axis labelled as $z$ such that the background magnetisation $m_z^{\mathrm{b}}$ (indicated by the black line in Fig.~\ref{fig:fig1}) generally divides the $S^2$ target space into two subspaces~\cite{Wu2022,Ackerman17, Samoilenka2017, Tai2018}. In this case, the base space changes from $S^3$ to $S^2\times S^1$\cite{Auckly2005, Hietarinta2009, Zheng2023, Kobayashi2013, BALAKRISHNAN2023, Knapman2024}. 
Linking numbers must be introduced for the top and bottom target subspaces individually (self-linking), as well as additional inter-linking numbers involving both subspaces.
Here the linking numbers and the Hopf index are calculated per period~\cite{Berger1984}.
Furthermore, $H$ is maintained as a measure of the Hopf index \textit{relative} to the unlinked state that has the same configuration at the boundary of the volume \cite{Berger1984, Prior2014}.

Magnetic textures with non-integer Hopf indices may be interpreted as states of ``mixed topology". Upon continuously tuning the background magnetisation $m_z^{\mathrm{b}}$ up or down the magnetisation texture can continuously evolve and the Hopf index is continuously modified without changing the linking numbers. Once the background magnetisation $m_z^{\mathrm{b}}$ reduces to a single point on $S^2$ the Hopf index becomes integer.
We illustrate this finding with a plethora of examples of smooth topological magnetic structures.
This work provides a unique framework to characterise, explain and classify the wealth of 3D topological textures in magnetism.

\subsection{Hopf index in terms of flux tubes}
\begin{figure}[tbp]
\includegraphics[width=0.95\columnwidth]{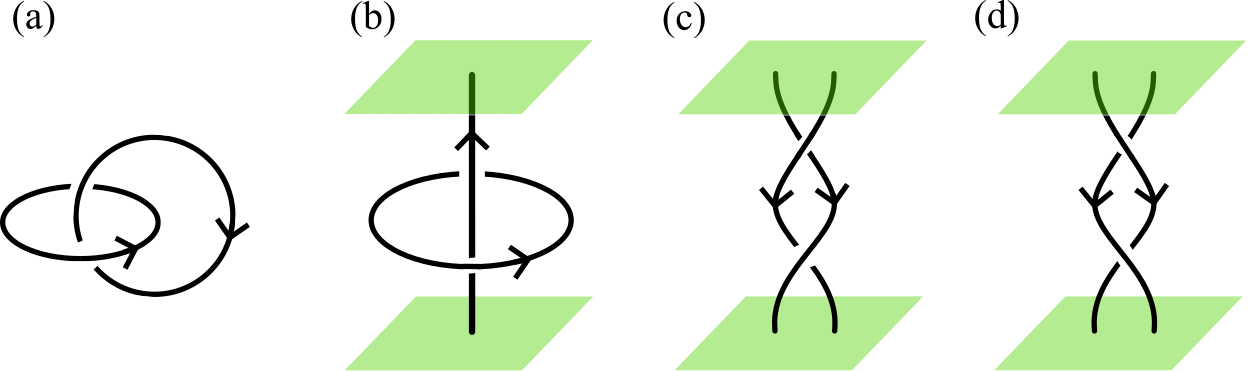}
	\caption{Here, we illustrate examples of two linked field lines. The green surfaces indicate periodic boundary conditions along the vertical axis.
    In panel (a), the field lines form closed, linked loops. Panel (b) shows a loop threaded by a vertical line. Panels (c) and (d) display vertically braided lines. The linking number of the field lines is $-1$ in scenarios (a-c), and $+1$ in (d) according to Eq.~\eqref{eq:linkingnum}.
    }
	\label{fig:linkingnumber}
\end{figure}

The Hopf index, or the helicity integral $H$, characterizes the topological properties of the field lines in a vector field $\mathbf{F}$, such as their linking number~\cite{Moreau1961, Moffatt1969, moffatt1978, Moffatt1981, Berger1984}. 
Field lines of $\mathbf{F}$ are oriented curves that are tangent at every point to the direction of $\mathbf{F}$.
The Hopf index can be written in a discrete geometric form for flux tubes~\footnote{Please note that every solenoidal vector field can be decomposed into flux tubes. This decomposition is not necessarily unique, but all decompositions lead to the same result in Eq.~\eqref{eq:Hopf_decomposed}. For a definition of flux tubes we refer to Ref.~\cite{Prior2014}.}, i.e.\ bundles of field lines as~\cite{Moffatt1969, Moffat1992, Ricca1992, Scheeler2017}, 
\begin{equation}
    H=\sum_{i=1}^{N_{\Phi}} L_{ii}\Phi_i^2 + 2\sum_{\substack{i=1\\ j\neq i}}^{N_{\Phi}}L_{ij}\Phi_i\Phi_j.
    \label{eq:Hopf_decomposed}
\end{equation}
Here, the vector field $\mathbf{F}$ has been decomposed into $N_\Phi$ flux tubes. $\Phi_i$ is the (positive definite) flux of the flux tube labelled by the index $i$, i.e.\ $\Phi_i=|\int \mathrm{d}\mathbf{a} \cdot \mathbf{F}|$, where the integration is over a cross-sectional area of the flux tube. 
 The linking numbers $L_{ii}$ and $L_{ij}$ are respectively the oriented self-linking of the flux tube $i$ and the oriented inter-linking of flux tubes $i$ and $j$.  
 $L_{ij}$ is computed using the Gauss linkage formula
 \begin{equation}
    L_{ij}=-\frac{1}{4\pi}\oint_{C_i} \oint_{C_j} \frac{\mathbf{r}_i-\mathbf{r}_j}{|(\mathbf{r}_i-\mathbf{r}_j)|^3}\cdot(d\mathbf{r}_i\times d\mathbf{r}_j)
    \label{eq:linkingnum}
\end{equation}
using the contour $C_i$ ($C_j$) following an arbitrarily chosen field line belonging to flux tube $i$ ($j$)~\footnote{Equation \eqref{eq:linkingnum} contains integrals over closed curves. For field lines that run vertically along the $z$-direction, the periodic boundary conditions imply that the field lines close trivially at infinity. An alternative is to define ``winding numbers'' of open field lines as in Ref.~\cite{Prior2014}}. 
$L_{ii}$ is calculated analogously using two field lines within the same tube $i$.
The sign of the linking numbers $L_{ii}$ and $L_{ij}$ are determined by the direction of the field lines as depicted in Fig.~\ref{fig:linkingnumber}~\footnote{Note that, in general, linking numbers can be written as the sum of twist and writhe components~\cite{Moffatt2013,Machon2019,Copar2011,Scheeler2017}.}, with further examples in Sec.~\ref{supp:math} of the Supplementary Material.

Eq.~\eqref{eq:Hopf_decomposed} decomposes $H$ into two contributions that, while geometrically distinct, are topologically equivalent. 
It also indicates that when the fluxes are non-integer, $H$ can assume non-integer values.

 A significant advantage of Eq.~\eqref{eq:Hopf_decomposed} is its intuitiveness and ease of analytical calculation in comparison with the mathematically equivalent well-known Whitehead integral formula~\cite{whitehead1947,Moffatt1969} $H=-\int \mathrm{d}^3r \, \mathbf{F}\cdot\mathbf{A}$ with $\mathbf{F}=\boldsymbol{\nabla}\times\mathbf{A}$. The latter can typically only be calculated numerically and its evaluation requires a carefully chosen gauge \cite{Knapman2024a}.

\subsection{Hopf index in magnetism}

In magnetism, the vector field $\mathbf{F}$ for which the Hopf index $H$ is typically calculated is the emergent magnetic field defined by~\cite{Volovik1987,Everschor2014}
\begin{equation}
    F^k=\frac{1}{8\pi}\epsilon^{ijk}\mathbf{m}\cdot(\partial_i\mathbf{m}\times\partial_j\mathbf{m}).
    \label{eq:Bem}
\end{equation}
where $\epsilon^{ijk}$ is the Levi-Civita symbol, $i,j,k\in\{x,y,z\}$ and the normalised vector field $\mathbf{m}$ represents the magnetic texture.

To calculate the Hopf index for the emergent magnetic field, Eq.~\eqref{eq:Hopf_decomposed} can generally be applied after the division of field lines of $\mathbf{F}$ into flux tubes.
Each flux tube contributes a flux $\Phi=|N_{\mathrm{sk}}|$, where 
$N_{\mathrm{sk}}=\int \mathrm{d}\mathbf{a} \cdot \mathbf{F}$ is the Skyrmion number calculated for the cross-sectional area of the flux tube of the emergent magnetic field.
Importantly, a flux tube can be arbitrarily decomposed into its constituent parts without affecting the validity of Eq.~\eqref{eq:Hopf_decomposed}, as we exemplify explicitly for a Skyrmion and a Hopfion in 
Sec.~\ref{supp:fluxtubes} of the Supplementary Material.  

In magnetism, it is common to consider the preimages of magnetisation $\mathbf{m}$, which are lines with constant $\mathbf{m}$, to obtain information about the Hopf index.
The preimages of $\mathbf{m}$ 
coincide at a local level with the field lines of $\mathbf{F}$ only if $\boldsymbol{\nabla}\cdot\mathbf{F}=0$~\cite{Barnett_2023}, i.e.\ in particular in the absence of Bloch points. 
In contrast to preimages, however, only the field lines of $\mathbf{F}$ possess a distinct direction, which is indispensable for determining unambiguously the sign of the linking number and hence the proper calculation of the Hopf index $H$, as emphasised by the sketches shown in  Fig.~\ref{fig:linkingnumber}. 

\begin{figure*}[htbp]
\includegraphics[width=0.8\textwidth]{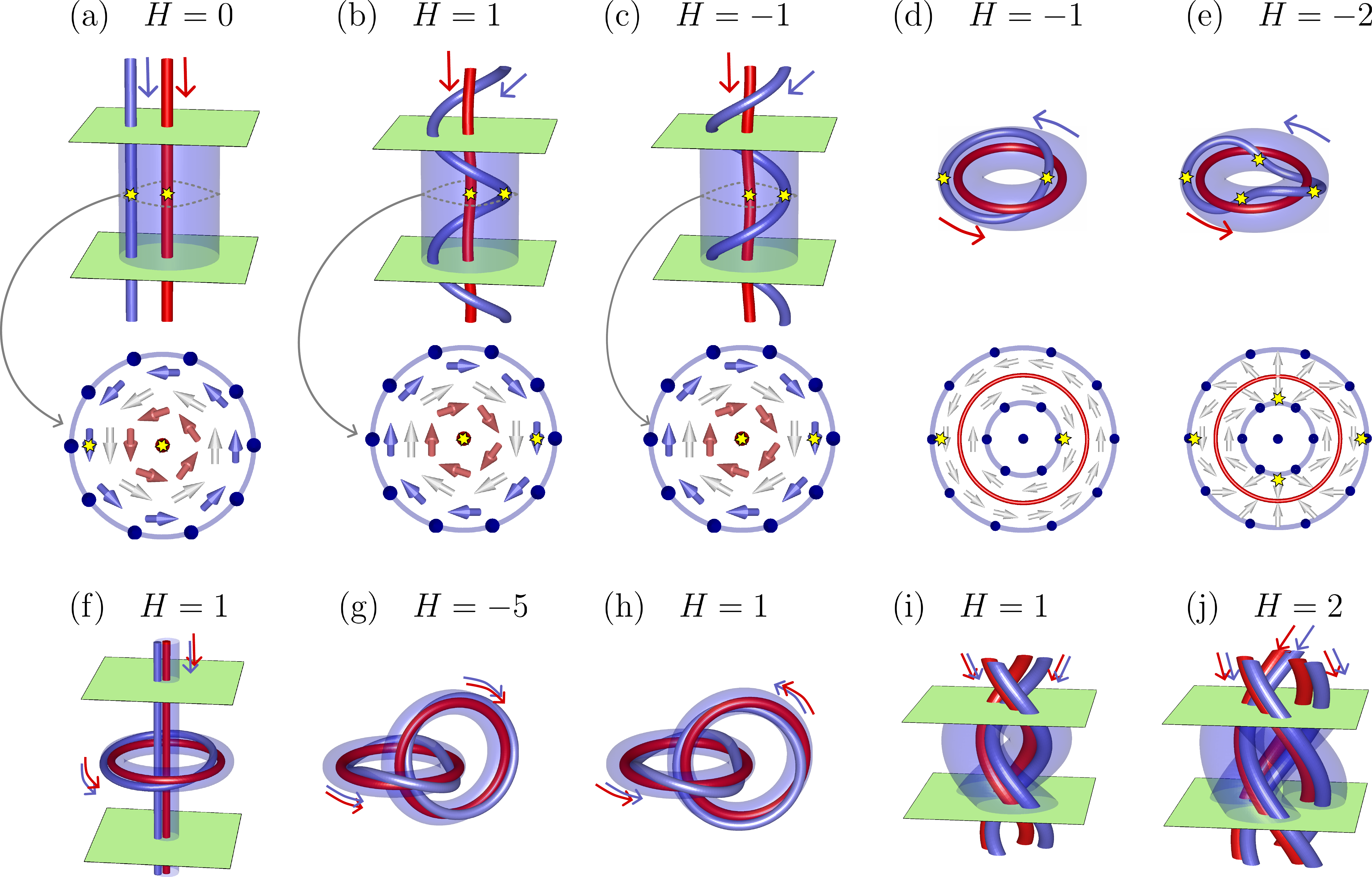}
	\caption{ \textbf{Illustrative examples applying Eq.~\eqref{eq:Hopf_decomposed} to spin textures with a ferromagnetic background ($m_z^{\mathrm{b}}=1$):}
	All panels show spin textures composed of $\Phi=1$ flux tubes enclosed by the transparent blue isosurface where $m_z=1$. For each flux tube two field lines of $\mathbf{F}$ are highlighted, which correspond to the preimages of $\mathbf{m}=(0,0,-1)$ and $\mathbf{m}=(0,-0.87,0.50)$, respectively accompanied by thin arrows showing the direction of the field line. The green planes represent the periodic boundary conditions along the vertical axis. The middle row shows cross-sections of the 3D structures depicted in a) - e) with the thick arrows indicating the spin direction. Yellow stars indicate the point of intersection of the field lines and the central horizontal plane. The colour code corresponds to the $S^2$ sphere shown in Fig.~\ref{fig:fig1}. 
    (a) A vertical Skyrmion string without self-linking,
    (b,c) vertical Skyrmion strings with self-linking $\pm 1$,
    (d) a Hopfion,
    (e) a Hopfion with self-linking $-2$,
    (f) a Hopfion encircling a vertical Skyrmion tube,
    (g,h) interlinked Hopfions,
    (i,j) braided Skyrmion tubes. 
}
	\label{fig:examplesfm}
\end{figure*}

\section{Analytical results}
As a central result, we demonstrate that the Hopf index is a continuously tunable function of the total magnetisation of the background. Notably, the value of $H$ can be adjusted through the application of external magnetic fields.

For periodic magnetic structures along $z$ with background $m_z^{\mathrm{b}}$ or those with a uniform background, we can express $H$ as
\begin{equation}
H =
\begin{cases} 
\in \mathbb{Z}
 & \text{if } |m_z^{\mathrm{b}}| = 1, \\
\in \mathbb{Z}/4
 & \text{if } m_z^{\mathrm{b}} = 0, \\
\in \mathbb{R}
 & \text{otherwise}.
\end{cases}
\label{eq:Hresult}
\end{equation}
While the result of the integer Hopf index can also be explained within the framework of standard homotopy group theory, the other results require more than homotopy group theory and can be proven with the help of Eq.~\eqref{eq:Hopf_decomposed}.

For magnetic textures embedded in a ferromagnet ($|m^b_z|=1$), the far-field maps to a single point on the $S^2$ sphere. 
In this case, the physical space can be compactified and $H$ needs to be necessarily integer-valued.

When the far field is not ferromagnetic and the background maps to a $S^1$ line on the $S^2$ sphere (see Fig.~\ref{fig:fig1}), such as for helimagnets~\cite{Leonov_2016,Rybakov2015, Voinescu2020,Kuchkin2023, Azhar2022}, $H$ is not guaranteed to be an integer. Instead, the Hopf index $H$ is a real number and depends on the total magnetisation of the background $m_z^{\mathrm{b}}$.
In this case, the noncollinear background splits the magnetisation $S^2$ target space into two topologically distinct subspaces. 
These subspaces map from specific loci in physical space (where flux tubes reside), enabling an unambiguous assignment of each flux tube to either of the two subspaces. This  necessitates the introduction of the concept of self-linking numbers for flux tubes belonging either to the top or bottom subspaces ($L_{11}$ and $L_{22}$), and the inter-linking number for two flux tubes belonging to different subspaces, $L_{12}$.

A special case arises for magnetic textures with a far-field described by $m^b_z=0$, where the $S^2$ sphere is split into half. In this case, the Skyrmion number of individual flux tubes is restricted to assume integer or half-integer values, i.e.\ $N_{\mathrm{sk}}\in \mathbb{Z}/2$. Since for periodic structures, it is always possible to find a decomposition of field lines into flux tubes where all linking numbers are integers, it follows from Eq.~\eqref{eq:Hopf_decomposed} that $H$ generally takes fractional values $H\in\mathbb{Z}/4$.

To demonstrate the versatility of Eqs.~\eqref{eq:Hopf_decomposed} and~\eqref{eq:Hresult} we now discuss specific examples.

Magnetic textures for which the flux tubes are not linked have a Hopf index of zero. Examples of this are magnetic structures translationally invariant along the $z$ direction such as vertical Skyrmion tubes (see Fig.~\ref{fig:examplesfm}(a)) or meron tubes (see Fig.~\ref{fig:spiralbackground}(a)).

In Sections~\ref{section:mzb1},~\ref{section:mzb0} and~\ref{section:mzbneq0} we discuss textures consisting of either a single flux tube, or multiple flux tubes spatially separated by the isosurface defined by $m_z^{\mathrm{b}}$, categorized according to Eq.~\eqref{eq:Hresult}. We summarize the calculations of key topological properties including the Hopf index of the corresponding magnetic textures in Tables~\ref{tab:fm} and~\ref{tab:spiral}.

\subsection{3D textures in ferromagnetic background $|m_z^{\mathrm{b}}|=1$}
\label{section:mzb1}

Fig.~\ref{fig:examplesfm} shows topological field configurations with a ferromagnetic background. Here the vector field can be decomposed into flux tubes which each have $\Phi=1$.

We further show two field lines for each flux tube, whose linking we determine according to the convention in Fig.~\ref{fig:linkingnumber}. Then we  compute the corresponding Hopf indices based on Eq.~\eqref{eq:Hopf_decomposed}.
When there is only a single flux tube, as in Fig.~\ref{fig:examplesfm} (a-e) then the equation for the Hopf index reduces to $ H=L_{11}\Phi_1^2=L_{11}$. For the examples shown in Fig.~\ref{fig:examplesfm}  (f-j) also the interlinking between pairs of tubes needs to be considered.
Note each Skyrmion tube that is inserted inside a Hopfion changes the Hopf index by $L_{ii}+2$ where $L_{ii}$ is the self-linking of the Skyrmion.
The calculation of the Hopf indices is summarized in Table~\ref{tab:fm}.

The examples shown, indicate that the classification in terms of linking numbers provides a more thorough description of magnetic textures than classifying them by topological $H$ and $N_{\textrm{sk}}$ indices only. For example, the two different spin textures are shown in Fig.~\ref{fig:examplesfm}(b) and (f) obey the same topological indices but different linking numbers.

\begin{table}[tbp]
    \centering
    \begin{tabular}{c}
   \includegraphics[width=1\columnwidth]{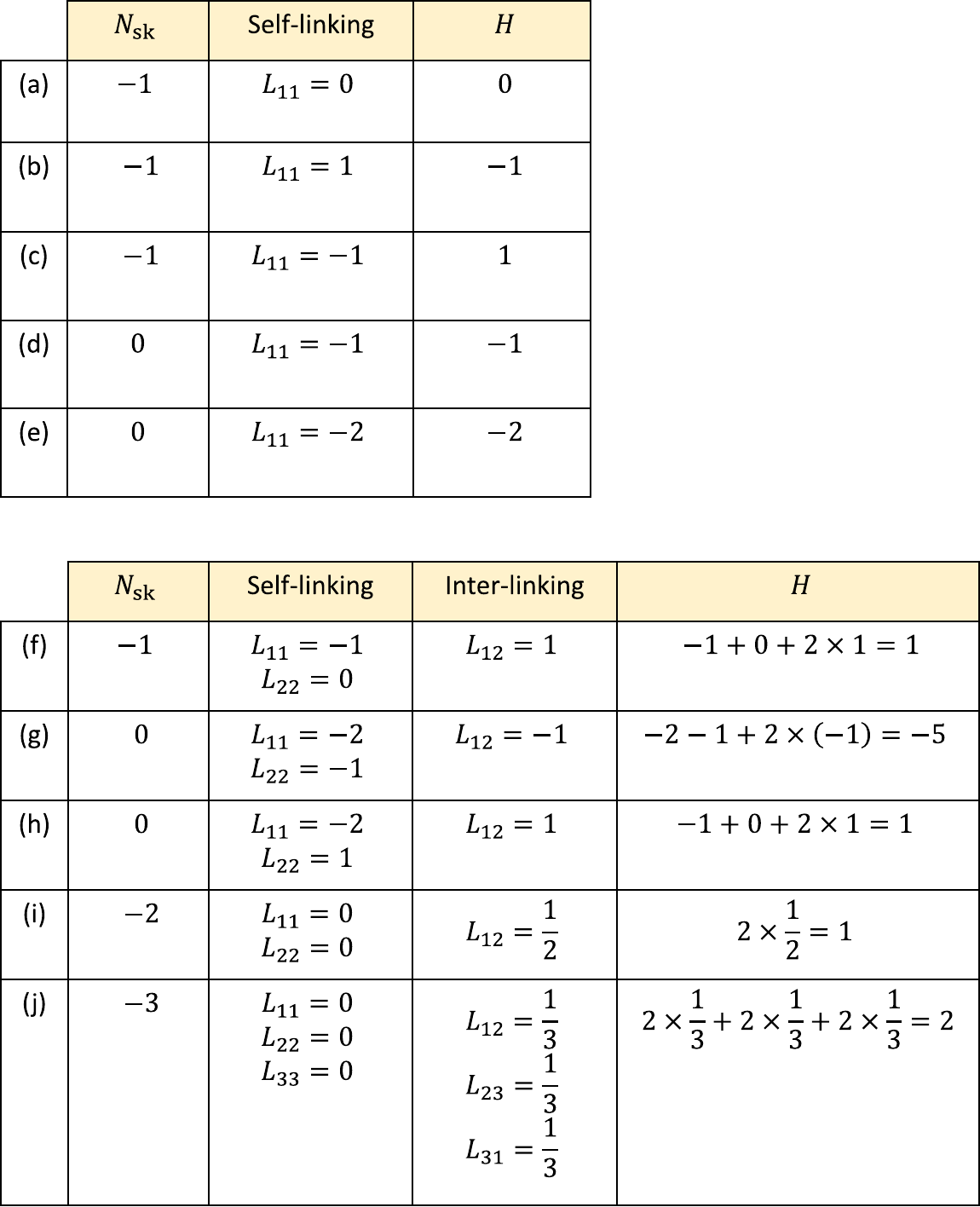}
    \end{tabular}
 \caption{Hopf index calculation for the spin textures shown in Fig.~\ref{fig:examplesfm}. The value of all fluxes is $\Phi_i=1$.
 }
\label{tab:fm}
\end{table}

\subsection{3D textures with $|m_z^{\mathrm{b}}|=0$:}
\label{section:mzb0}

\begin{figure*}[htbp]
\includegraphics[width=0.85\textwidth]{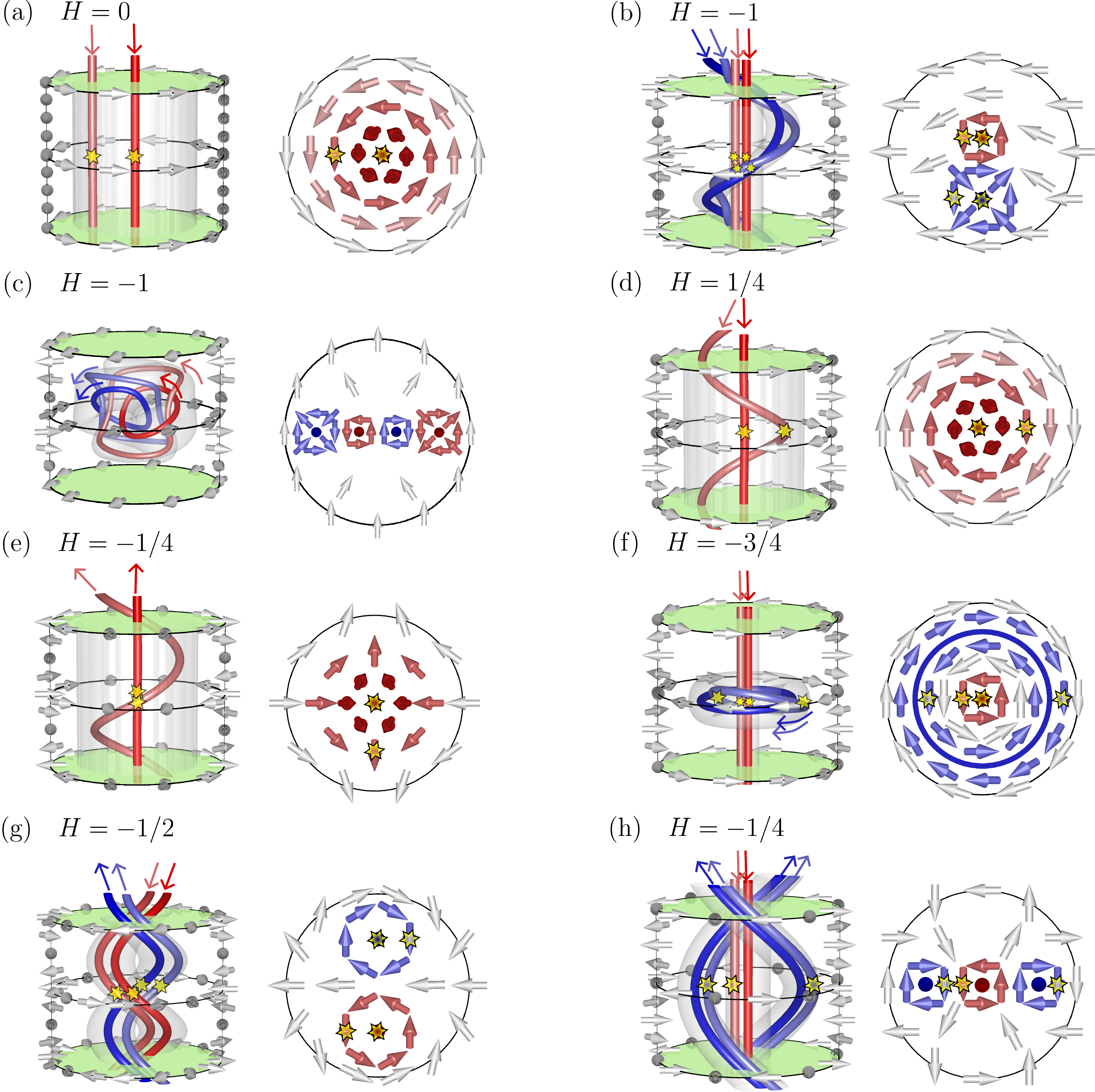}
	\caption{ \textbf{Illustrative examples applying Eq.~\eqref{eq:Hopf_decomposed} to spin textures with a spiral background ($m_z^{\mathrm{b}}=0$):} 
 All panels show spin textures composed of $\Phi=1/2$ flux tubes enclosed by the transparent white isosurface where $m_z=0$.
 For each flux tube, two field lines of $\mathbf{F}$ are highlighted, which correspond to the preimages of $\mathbf{m}=(0,0,\pm 1)$ and $\mathbf{m}=(0,-0.87,\pm 0.50)$, respectively accompanied by thin arrows showing the direction of the field line. 
 The green planes represent the periodic boundary conditions along the vertical axis. The second panel shows cross-sections of the 3D structures through the middle plane, with yellow stars indicating the point of intersection of the field lines and this plane, and thick arrows indicating the direction of $\mathbf{m}$. The colour code corresponds to the $S^2$ sphere shown in Fig.~\ref{fig:fig1}. We depict (a) a meron (b) Skyrmion and (c) Heliknoton, in a spiral background. (d-e) meron with self-linking  (f) 
 Twiston (a vertical meron tube surrounded by a self-linked meron torus) (g) two braided meron tubes (h) three braided meron tubes. 
	}
	\label{fig:spiralbackground}
\end{figure*}

When the far field of the texture is an $S^1$ line, the $S^2$ sphere is naturally split into two subspaces. The corresponding magnetic textures can be decomposed into 
fractional components.
For $|m_z^{\mathrm{b}}|=0$ a distribution of space into flux tubes can be made such that all flux tubes carry a flux of $\Phi=\frac{1}{2}$.
Examples of such textures are illustrated in Fig.~\ref{fig:spiralbackground}.

To construct an $S^1$ background, we describe the magnetisation using the angular parameterization 
\begin{equation}
    \bm{m}=(\cos\theta\cos\phi,\cos\theta\sin\phi,\sin\theta).
    \label{eq:m}
\end{equation}
In the far field, the spins can either rotate along the $z$-axis (as a helical spiral $\phi=qz$ with ferromagnetic ordering within each $xy$-plane, see Fig.~\ref{fig:spiralbackground}(b,c)) or assume an in-plane vorticity, 
($\phi=\nu\chi$, where $\nu$ is an integer and $\chi=\arctan(y/x)$, while being invariant along the $z$-direction, see Fig.~\ref{fig:spiralbackground}(a)), or a combination of both a spiral and a vortex state, 
which is a screw dislocation~\cite{Azhar2022} ($\phi=\nu\chi+qz$, where here $\nu$ is the integer quantising the Burgers vector, see Fig.~\ref{fig:spiralbackground}(d-g)). In the following sections, we describe these textures in detail and apply Eq.~\eqref{eq:Hopf_decomposed} to them.

\subsubsection*{(a) Meron}
A meron is a texture characterized by a vortex-like far-field ($\phi=\chi$), and a single vertical flux tube at its core, as illustrated in Fig.~\ref{fig:spiralbackground}(a). The magnetisation maps to half of the $S^2$ sphere, resulting in a flux value of $\Phi=1/2$. The flux tube shown is aligned along the $z$-axis, with all field lines of $\bm{F}$ running parallel to the $z$-axis. 
Since the flux tube of the magnetic texture is not linked, the Hopf index is zero, as discussed above.

\subsubsection*{(b) Skyrmion in spiral background}
 Fig.~\ref{fig:spiralbackground}(b) shows a Skyrmion with $N_{\mathrm{sk}}=-1$ embedded in a spiral background. The Skyrmion string is aligned parallel to the $z$-axis, which is the direction of the wavevector of the spiral. Each $xy$-plane contains a vortex and an antivortex with opposite core magnetisation $m_z$.
Analysing the linkage of the field lines of $\mathbf{F}$, the vortex tube has a self-linking number $L_{11}=0$, while the antivortex tube has $L_{22}=-2$. The inter-linking between these two tubes is $L_{12}=-1$. Thus, we obtain a Hopf index of $H=L_{11}\Phi_1^2+L_{22}\Phi_2^2+2L_{12}\Phi_1\Phi_2=-1$.


\subsubsection*{(c) Heliknoton}
A Heliknoton \cite{Voinescu2020} can be interpreted as a Hopfion embedded within a spiral background, see Fig.~\ref{fig:spiralbackground}(c). It consists of two flux tubes mapping to the upper and lower half of the $S^2$ sphere respectively, each with a flux value of $\Phi_1=\Phi_2=1/2$ and linking numbers $L_{11}=L_{22}=L_{12}=-1$. 
The Skyrmion number in any cross-sectional plane is given by $N_{\mathrm{sk}}=0$ and the Hopf index evaluates to $H=-1$.

 
\subsubsection*{(d-e) Vortex and antivortex tubes with self-linking}

Fig.~\ref{fig:spiralbackground}(d) and (e) show special types of screw dislocations composed of a single flux tube with self-linking and a Burgers integer of $|\nu|=1$\footnote{(referred to as sd$_{1}^{+}$ and sd$_{-1}^{+}$ in Ref.~\cite{Azhar2022})}. In (d) and (e) the cross-section is a vortex ($\nu=1$) and an anti-vortex ($\nu=-1$), respectively.

For such single flux tube screw dislocations with Burgers integer~$\nu$ the flux is $\Phi=|\nu/2|$ and the self-linking is $L_{11}=1/\nu$. The Skyrmion charge is given by $N_{\mathrm{sk}}=p\nu/2$, where $p$ denotes the core polarity (the sign of $m_z$ at the core) and the Hopf index, is $H=L_{11}\Phi_1^2=\nu/4$.

\subsubsection*{(f) Twiston}
Fig.~\ref{fig:spiralbackground}(f) shows a special type of screw dislocation with $\nu=1$ which we denote as ``Twiston'' in the following~\footnote{(referred to as sd$_{1}^{\mathrm{Sk-}}$ in Ref.~\cite{Azhar2022})}. It is composed of two flux tubes: at the centre of the structure is a vertical meron tube with parameters $\Phi_1=1/2$, $L_{11}=0$, and polarity $p$. 
Surrounding this is a meron-torus with $\Phi_2=1/2$, $L_{22}=-1$, and opposite polarity $-p$. 
There is inter-linking between these tubes, with a linking number $L_{12}=-1$.
The Skyrmion number for each cross-section is $N_{\mathrm{sk}}=p/2$
and the Hopf index $H$ for this configuration is $H=-3/4$.

\subsubsection*{ (g) Screw dislocation with $\nu=2$}
The screw dislocation illustrated in Fig.~\ref{fig:spiralbackground}(g) consists of two flux tubes that do not have any self-linking, i.e., $L_{11}=L_{22}=0$. Each flux tube has a flux $\Phi_1=\Phi_2=1/2$, and their inter-linking is $L_{12}=-1$. The Skyrmion charge in the $xy$-plane for this configuration is $N_{\mathrm{sk}}=0$. The Hopf index is $H=-1/2$.

\subsubsection*{(h) Screw dislocation with $\nu=3$}
Fig.~\ref{fig:spiralbackground}(h) shows a screw dislocation with $\nu=3$, which consists of three meron flux tubes each with flux $\Phi_1=\Phi_2= \Phi_3 = 1/2$. None of the flux tubes have self-linking, i.e., $L_{11}=L_{22}=L_{33}=0$, however their inter-linking numbers are $L_{23}=1/2$ and $L_{12}=L_{13}=-1/2$. The Skyrmion charge in the $xy$-plane for this configuration, contributed by the three merons, is $N_{\mathrm{sk}}=1/2$. The Hopf index is $H=-1/4$.

\begin{table*}[]
    \centering
    \begin{tabular}{c}
                \includegraphics[width=0.75\textwidth]{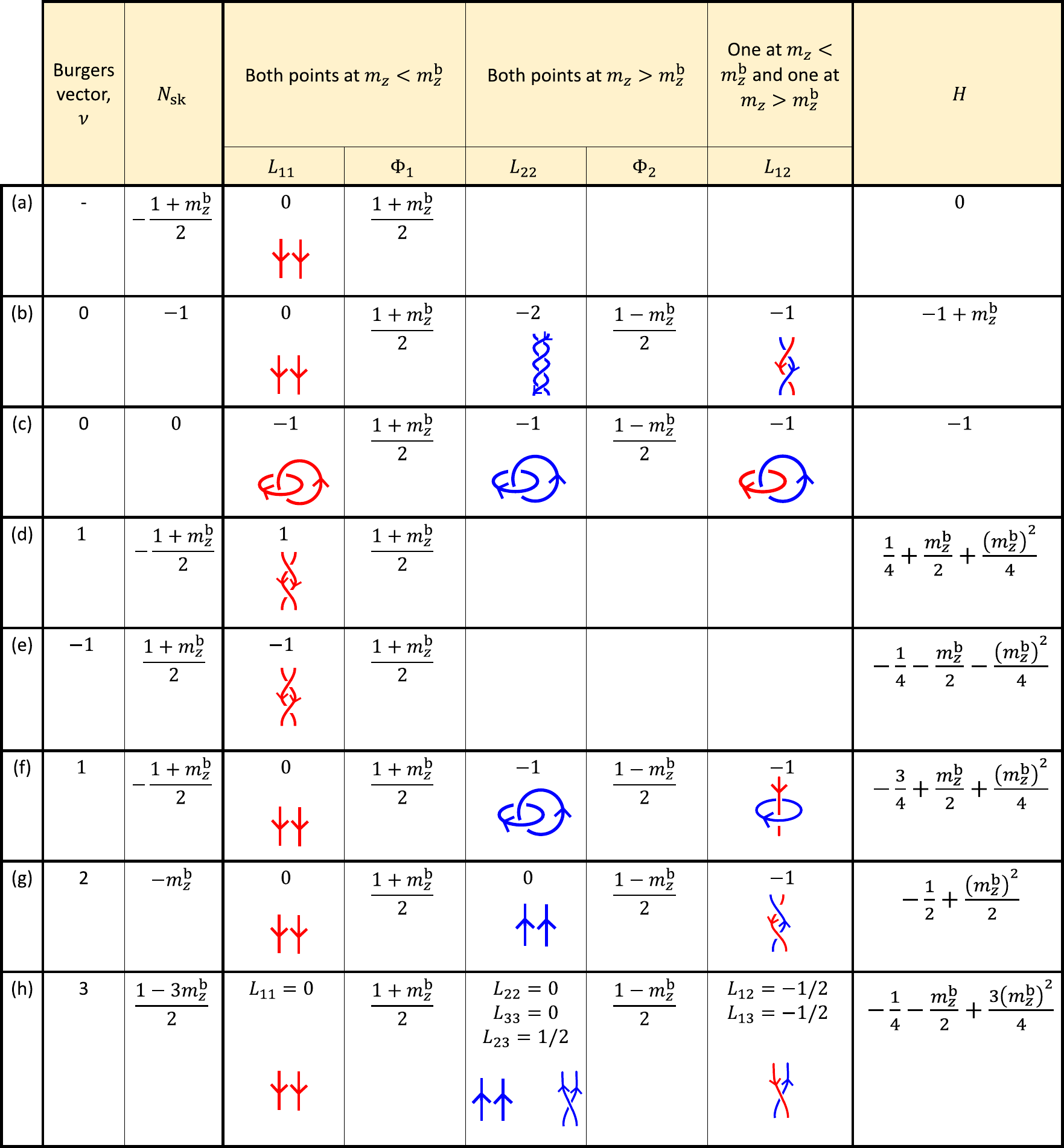}
    \end{tabular}
	\caption{ Calculation of the Hopf index for the textures shown in Fig.~\ref{fig:spiralbackground} with general background magnetisation $m_z^\mathrm{b}$.}
	\label{tab:spiral}
\end{table*}

\subsection{3D textures with $0<|m_z^{\mathrm{b}}|<1$}
\label{section:mzbneq0}

For magnetic textures where the background magnetisation is neither along the $z$ direction nor in the $xy$-plane, the results obtained in Sec.~\ref{section:mzb0} can be generalized.

For a background magnetisation with $0<|m_z^{\mathrm{b}}|<1$ the $S^1$ line on the $S^2$ sphere shifts in latitude (see Fig.~\ref{fig:fig1}). This enables the partitioning of physical space into flux tubes, such that all flux tubes corresponding to the upper and lower subspaces on the $S^2$ sphere possess flux values of $(1-m_z^{\mathrm{b}})/2$ and  $(1+m_z^{\mathrm{b}})/2$ respectively. The linking numbers, however, remain invariant under variations of $m_z^{\mathrm{b}}$, similar to examples from fluid dynamics~\cite{Moffatt2013}.
In general, $N_\mathrm{sk}$ \cite{delSer2024} and $H$ become a function of $m_z^{\mathrm{b}}$ and are continuously tuneable as shown in Table.~\ref{tab:spiral}.

Changing the magnetisation background continuously transforms the magnetic structures considered in Fig.~\ref{fig:spiralbackground}. Importantly for $0<|m_z^{\mathrm{b}}|<1$, the magnetic textures are in a mixed topology state and can transform smoothly to states with integer Hopf indices, which can be different for tuning the background magnetisation up ($m_z^{\mathrm{b}} = 1$) or down ($m_z^{\mathrm{b}} =-1$).

For example, the Skyrmion in a spiral background transforms 
for $m_z^{\mathrm{b}} = 1$ into a Skyrmion in a ferromagnetic background with $H=0$. Conversely, for $m_z^{\mathrm{b}} = -1$ it becomes an anti-Skyrmion with a self-linking. 

The Heliknoton is special in the sense that its topological indices $N_{\mathrm{sk}}$ and $H$ do not depend on $m_z^{\mathrm{b}}$. Aligning the background continuously along $z$ the structure transforms to a Hopfion in a ferromagnetic background, i.e.\ a single closed flux tube as in Fig.~\ref{fig:examplesfm}(d).
The (anti-)vortex with self-linking (Fig.~\ref{fig:spiralbackground}(d,e) transforms for $m_z^{\mathrm{b}}= -1$ to a ferromagnetic state and for $m_z^{\mathrm{b}}= 1$ to a (anti-)Skyrmion with self-linking.
The Twiston transforms for $m_z^{\mathrm{b}}= -1$ to a Hopfion with $H=-1$ and for $m_z^{\mathrm{b}}= 1$ to a Skyrmion with $H=0$ as shown in Fig.~\ref{fig:examplesfm}(a). Thus, this texture with $(N_{\mathrm{sk}},H)=(-1/2,-3/4)$ at $m_z^{\mathrm{b}}=0$ can be converted into either a Skyrmion tube or a Hopfion by
changing the background (e.g.\ by an external magnetic field) to a ferromagnetic state. 
The screw dislocation with $\nu=2$ transforms for $m_z^{\mathrm{b}}= \pm1$ to a Skyrmion with $(N_{\mathrm{sk}},H)=(\mp1,0)$.
The screw dislocation with $\nu=3$ transforms for $m_z^{\mathrm{b}}=1$ to a Skyrmion with $(N_{\mathrm{sk}},H)=(-1,0)$, and for  $m_z^{\mathrm{b}}=-1$ to a pair of braided Skyrmion tubes with $(N_{\mathrm{sk}},H)=(2,1)$.

\section{Numerical Results}

\begin{figure*}
	\includegraphics[width=0.85\textwidth]{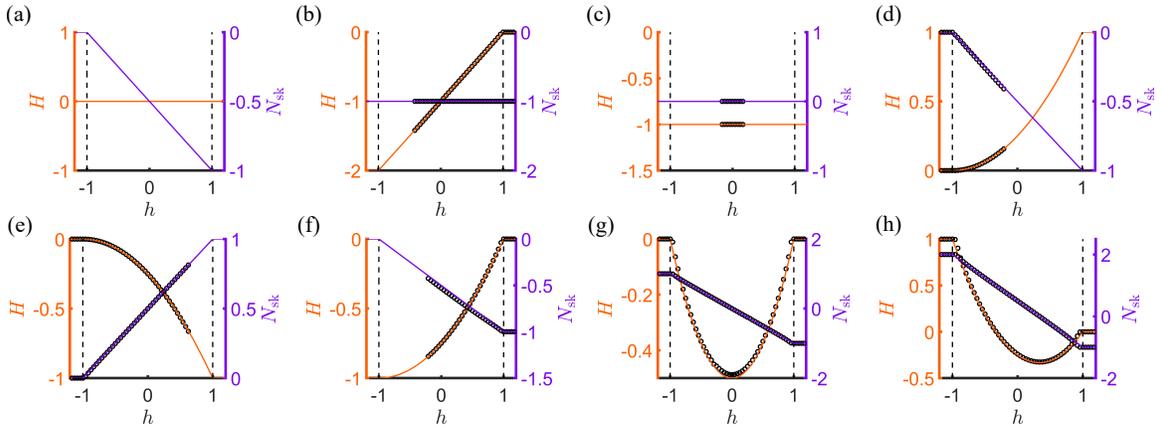}
	\caption{Hopf index $H$ and Skyrmion number $N_{\mathrm{sk}}$ for the magnetic textures depicted in Fig.~\ref{fig:spiralbackground}, described by the analytical expressions in Table~\ref{tab:spiral} (lines). In (b-g) the black circles are the numerically calculated values of $N_{\mathrm{sk}}$ and $H$ for the model of Eq.~\eqref{eq:chiralMagnets}, for the range of $|h|<1.2$, where the texture was numerically stable.
}
	\label{fig:Numerics_spiralbackground}
 \end{figure*}

 We employ a specific chiral magnet model to stabilise the examples of topological defects embedded within a non-uniform state, that were considered in the previous section. We emphasise, however, that the methods presented for the calculation of $H$ using Eq.~\eqref{eq:Hopf_decomposed} are model-independent. 
 
 We consider the energy functional~\cite{Dzyaloshinkskii1964}
\begin{equation}
    E=\int \mathrm{d}^3r \, [A\, \partial_i\mathbf{m}\cdot\partial_i\mathbf{m} +D \,\mathbf{m}\cdot\boldsymbol{\nabla}\times\mathbf{m} -\mu_0M_{\mathrm{s}}\, \mathbf{H}\cdot\mathbf{m}
    ],
\label{eq:chiralMagnets}
\end{equation}
where $A$ is the ferromagnetic exchange stiffness, $D$ is the Dzyaloshinskii-Moriya interaction (DMI) strength, 
$M_{\mathrm{s}}$ is the saturation magnetisation and $\mathbf{H}$ represents the external magnetic field.

 To simplify the analysis, we introduce 
 a dimensionless magnetic field $h=|\mathbf{H}|/H_{\mathrm{c2}}$ measured in units of $H_{\mathrm{c2}}=D^2/(2A\mu_0M_{\mathrm{s}})$. 
For an applied magnetic field along the $z$ direction, the ground state of Eq.~\eqref{eq:chiralMagnets} is either a ferromagnetic (for $|h|\geq1$) or a (conical) spiral (for $|h|<1$) state.
 The spiral state is characterized by a 
wavevector along $z$ with magnitude $q=4\pi A/D$ and
\begin{equation}
 \theta=\arccos(h),  \quad   \phi=qz,
\end{equation}
in the parametrization of the magnetisation given in Eq.~\eqref{eq:m}.

In addition, for $|h|<1$ screw dislocations with Burgers vector $\nu$ are also energetically metastable solutions of Eq.~\eqref{eq:chiralMagnets}~\cite{Azhar2022}, with the far field given by,
\begin{equation}
 \theta=\arccos(h), \quad    \phi=\nu\chi+qz.
\end{equation}
In both cases, conical spiral and screw dislocation, the background magnetisation is determined by $m_z^{\mathrm{b}}=\cos{\theta}=h$.

We conduct micromagnetic simulations based on the model in Eq.~\eqref{eq:chiralMagnets} and relax the textures shown in Fig.~\ref{fig:spiralbackground} for a range of rescaled magnetic field values $-1.2<h<1.2$.
Simulations were performed using \textsc{MuMax3}~\cite{Vansteenkiste2014} with the following parameters: exchange stiffness $A = \SI{4e-13}{\joule\per\meter}$, bulk DMI constant $D = \SI{2.8e-4}{\joule\per\square\meter}$, and saturation magnetisation $M_{\mathrm{s}} = \SI{1.63e5}{\ampere\per\meter}$. The grid size was set to $(N_x,N_y,N_z) = (360,360,36)$, with a cell size of $\SI{0.5}{\nano\meter}$, and a cylindrical geometry with a diameter of $\SI{180}{\nano\meter}$. 
Periodic boundary conditions were applied along the $z$-axis, while open boundary conditions were used along the $x$ and $y$ axis. Using these parameters, we obtain $L_{\mathrm{D}}=4\pi A/D=\SI{18}{\nano\meter}$ as the period of the spiral along the $z$-direction, and $H_{\mathrm{c2}}=\SI{0.6}{\milli\tesla}$ as the critical field for transition to the ferromagnetic state.

Note that depending on the model considered, the textures deform from the ideal textures shown in Fig.~\ref{fig:spiralbackground} 
upon relaxation, without changing their topological characteristics, see Sec.~\ref{supp:numerics} of the 
Supplementary Material.

The numerically calculated values of $N_{\mathrm{sk}}$ and $H$ are presented in Fig.~\ref{fig:Numerics_spiralbackground} (shown as circles) alongside the analytical results (lines) from Table~\ref{tab:spiral}, showing good agreement (details about the calculation of $H$ are presented in Sec. \ref{supp:numerics} of the Supplementary Material). The numerical results are shown for the range of magnetic field values (varied in steps of $\SI{0.025}{\milli\tesla}$) where we found the texture to be numerically (meta)stable in this specific model with the numerical implementation chosen. Notably, the analytical expressions in Table~\ref{tab:spiral} allow us to predict the resulting textures for $|m_z^\mathrm{b}|=1$, even when these textures are not energetically stable within the model considered.

The model Eq.~\eqref{eq:chiralMagnets} is invariant under the following simultaneous transformation of magnetisation and magnetic field: $\mathbf{m}\rightarrow-\mathbf{m}$, $\mathbf{H}\rightarrow-\mathbf{H}$. This means that all magnetisation configurations exist with two different polarities, which actually have opposite  Skyrmion numbers while having the same Hopf index, i.e.\ $N_{\mathrm{sk}}\rightarrow -N_{\mathrm{sk}}$ and $H\rightarrow H$ under this transformation.

\hfill \break

 \section{Conclusion}
To conclude, we introduced and implemented a discrete formulation of the Hopf index (Eq.~\eqref{eq:Hopf_decomposed}) that provides a more intuitive understanding as an alternative to the commonly used Whitehead integral formula.
This approach groups the emergent magnetic field into flux tubes with distinct linking numbers, providing clearer insights into the topological characterization of magnetic textures. While the Whitehead integral can only be computed analytically for a very limited set of ansatzes, the discrete formulation (Eq.~\eqref{eq:Hopf_decomposed}) provides precise and model-independent analytical results once the flux tubes and linking numbers are identified. 

Thereby we have shown that non-integer values of the Hopf index naturally occur for magnetic textures with a non-uniform background magnetisation.
Importantly magnetic textures in a ``mixed-topology'' state --- such as the Twiston --- can transform smoothly into different topological sectors upon continuously tuning the background magnetisation. While both the Skyrmion number and the Hopf index are continuous functions of the background magnetisation, the linking numbers remain conserved topological invariants, highlighting their significance as topological characteristics of the texture.
Our results establish a strong physical foundation for the existence of magnetic textures with arbitrary non-integer Hopf indices.

 \section{Acknowledgements}
We thank Markus Garst, Robin Msiska, Finn Feldkamp, Stefan Bl\"ugel, Nikolai Kiselev, Jan Masell, Volodymyr Kravchuk and Sopheak Sorn for helpful discussions. This work was supported by the German Research Foundation (DFG) Project No.~505561633 in the TOROID project co-funded by the French National Research Agency (ANR) under Contract No.~ANR-22-CE92-0032. We further acknowledge funding from the German
Research Foundation (DFG) Project No.~320163632 (Emmy
Noether), Project No.~403233384 (SPP2137 Skyrmionics) and Project No. 278162697-SFB 1242 (project B10).

\newpage

\appendix

\section{Linking numbers}
\label{supp:math}

\subsection{Linking numbers}
\begin{figure}
\includegraphics[width=0.9\columnwidth]{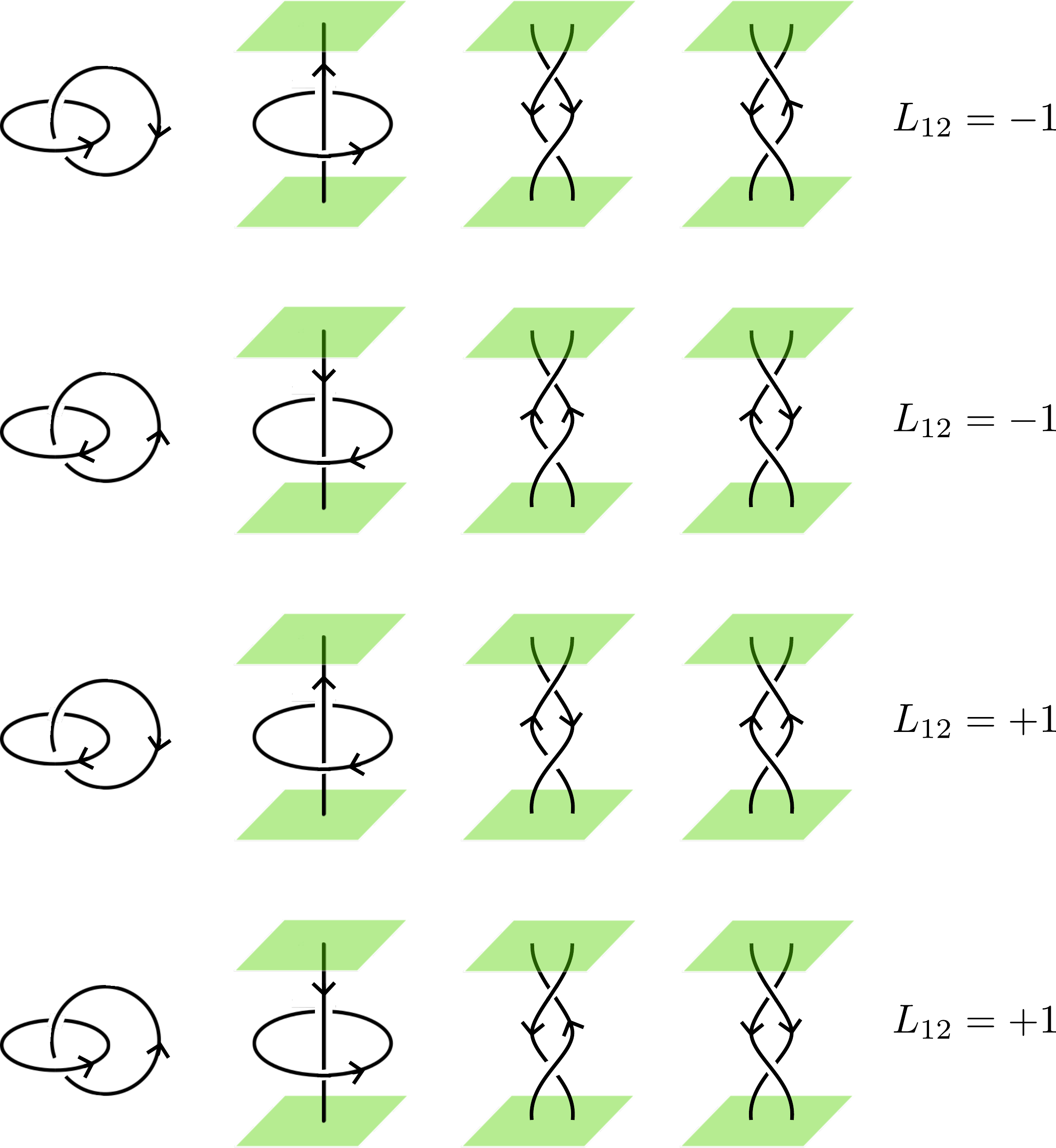}
	\caption{The values of the linking numbers $L_{12}$ for two oriented curves, calculated using the formula given in Eq.~\eqref{eq:linkingnum}.  $L_{12}=-1$ for the first two rows and  $L_{12}=1$ for the last two rows.}
 \label{fig:linking_supplementary}
\end{figure}

The convention for linking numbers followed in this paper is given by the formula in Eq.~\eqref{eq:linkingnum}, and sketched in Fig.~\ref{fig:linkingnumber} of the main text. Reversing the direction on any contour reverses the sign of the linking number, as shown in Fig.~\ref{fig:linking_supplementary}.

\subsection{Whitehead formula in terms of linking numbers}
In this section, it will be shown that the Whitehead formula has an intuitive interpretation related to the Gauss linking number, using the Coulomb gauge, $\mathbf{\nabla}\cdot\mathbf{A}=0$.

Using the Whitehead formula,
\begin{equation}
    H=-\int \mathrm{d}^3r \, \mathbf{F}\cdot\mathbf{A}
\end{equation}
and the Fourier transform of the equations $\mathbf{\nabla}\times\mathbf{A}=\mathbf{F}$ $\mathbf{\nabla}\cdot\mathbf{A}=0$, we arrive at the result, 
\begin{equation}
    H=-\frac{1}{(2\pi)^3}\int \mathrm{d}^3k \, \frac{\mathbf{k}\cdot (\mathbf{F}(\mathbf{k}) \times \mathbf{F}(-\mathbf{k})) }{|\mathbf{k}|^2} .
\end{equation}
Substituting $\mathbf{F}(\mathbf{k})=\int \mathrm{d}^3r \, \mathbf{F}(\mathbf{r}) e^{-i \mathbf{k}\cdot\mathbf{r}}$, we obtain \cite{Kravchuk2023},
\begin{equation}
    H=-\frac{1}{4\pi}\int \mathrm{d}^3r \int \mathrm{d}^3 r^\prime \frac{(\mathbf{r}-\mathbf{r}^\prime)\cdot (\mathbf{F}(\mathbf{r})\times\mathbf{F}(\mathbf{r}^\prime))}{|\mathbf{r}-\mathbf{r}^\prime|^3}.
    \label{eq:H_coulomb}
\end{equation}

We now consider a closed curve $\mathbf{C}_i: [0,1] \mapsto \mathbb{R}^3 $, parametrized by $s \in [0,1]$, with $\mathbf{C}_i(0)=\mathbf{C}_i(1)$.
We define a vector field $\mathbf{V}: \mathbb{R}^3 \to \mathbb{R}^3$ such that $\mathbf{V}_i(\mathbf{r}) = \mathbf{0}$ for $\mathbf{r} \notin \mathbf{C}_i$, while along the curve $\mathbf{C}_i$ it satisfies

\begin{equation}
    \frac{d\mathbf{C}_i(s)}{ds}\times \mathbf{V}_i\left(\mathbf{C}_i(s)\right)=0,
\end{equation}
meaning that $\mathbf{C}_i(s)$ is a streamline for $\mathbf{V}_i$. The vector field $\mathbf{V}_i$ can therefore be written as

\begin{equation}
\mathbf{V}_i(\mathbf{r}) = \int_0^1 \mathrm{d}s \ \frac{d \mathbf{C}_i(s)}{ds}  \delta^{(3)}\left( \mathbf{r} - \mathbf{C}_i(s) \right),
\end{equation}
where  $\delta^{(3)}(\mathbf{r}-\mathbf{r}(s))$ is the three-dimensional Dirac delta function ensuring that $\mathbf{V}_i$ is localized at the contour $\mathbf{C}_i(s)$. 

For our purposes, we consider now two closed curves $\mathbf{C}_i(s)$ and $\mathbf{C}_j(s')$ according to the definitions above and we consider the two vector fields $\mathbf{V}_i$ and $\mathbf{V}_j$. 

The linking number between the curves $\mathbf{C}_i(s)$ and $\mathbf{C}_j(s')$ can be calculated by using \eqref{eq:H_coulomb} with the vector fields $\mathbf{V}_i$ and $\mathbf{V}_j$.

By direct substitution one gets

\begin{equation}
\begin{split}
    H &= -\frac{1}{4\pi} \int d^3 r \int d^3 r^\prime \frac{(\mathbf{r}-\mathbf{r}^\prime)}{|\mathbf{r}-\mathbf{r}^\prime|^3}\\
    & \cdot \int_0^1 ds \  \frac{d \mathbf{C}_i(s)}{ds}  \delta^{(3)}\left( \mathbf{r} - \mathbf{C}_i(s) \right) \\
    &\times \int_0^1 ds^\prime \ \frac{d \mathbf{C}_j(s^\prime)}{\mathrm{d}s^\prime}  \delta^{(3)}\left( \mathbf{r}^\prime - \mathbf{C}_j(s^\prime) \right).
\end{split}
\end{equation}

Now, using the fact that $\int_0^1 ds \ \frac{d \mathbf{C}_i(s)}{ds} = \oint_{\mathbf{C}_i} d \mathbf{r}_i$, after having integrated the delta functions contributions, one can arrive  at the result in  Eq. \eqref{eq:linkingnum} for the $H$ calculated for a pair of thin field lines, which simply equals $L_{ij}$:
\begin{equation}
    L_{ij}=-\frac{1}{4\pi}\oint_{C_i} \oint_{C_j} \frac{\mathbf{r}_i-\mathbf{r}_j}{|\mathbf{r}_i-\mathbf{r}_j|^3}\cdot(d\mathbf{r}_i\times d\mathbf{r}_j).
\end{equation}


Hence the expression in Eq.~\eqref{eq:H_coulomb} is analogous to Eq.~\eqref{eq:linkingnum}, and they become equivalent in the limit where the field $\mathbf{F}(\mathbf{r})$ is sharply localised on a contour. This correspondence supports the interpretation of the Hopf index as in Eq.~\eqref{eq:Hopf_decomposed}, as a flux-weighted measure of the linkage of field lines of $\mathbf{F}$. 
\\

\section{Division of vector field into flux tubes}
\label{supp:fluxtubes}
The assignment of $\mathbf{F}$ to flux tubes for an arbitrary spin texture is not unique, in general. We illustrate this with specific examples of a Skyrmion with self-linking (section \ref{sm:fluxdecomposition_sk}) and  a Hopfion (section \ref{sm:fluxdecomposition}) in a ferromagnetic background. In all cases, the boundaries of flux tubes must be surfaces with $\mathbf{F}\cdot\hat{n}|_S=0$, i.e. the field lines of $\mathbf{F}$ cannot enter or leave this surface.

\subsection{Flux division for one Skyrmion}
\begin{figure}
\includegraphics[width=1\columnwidth]{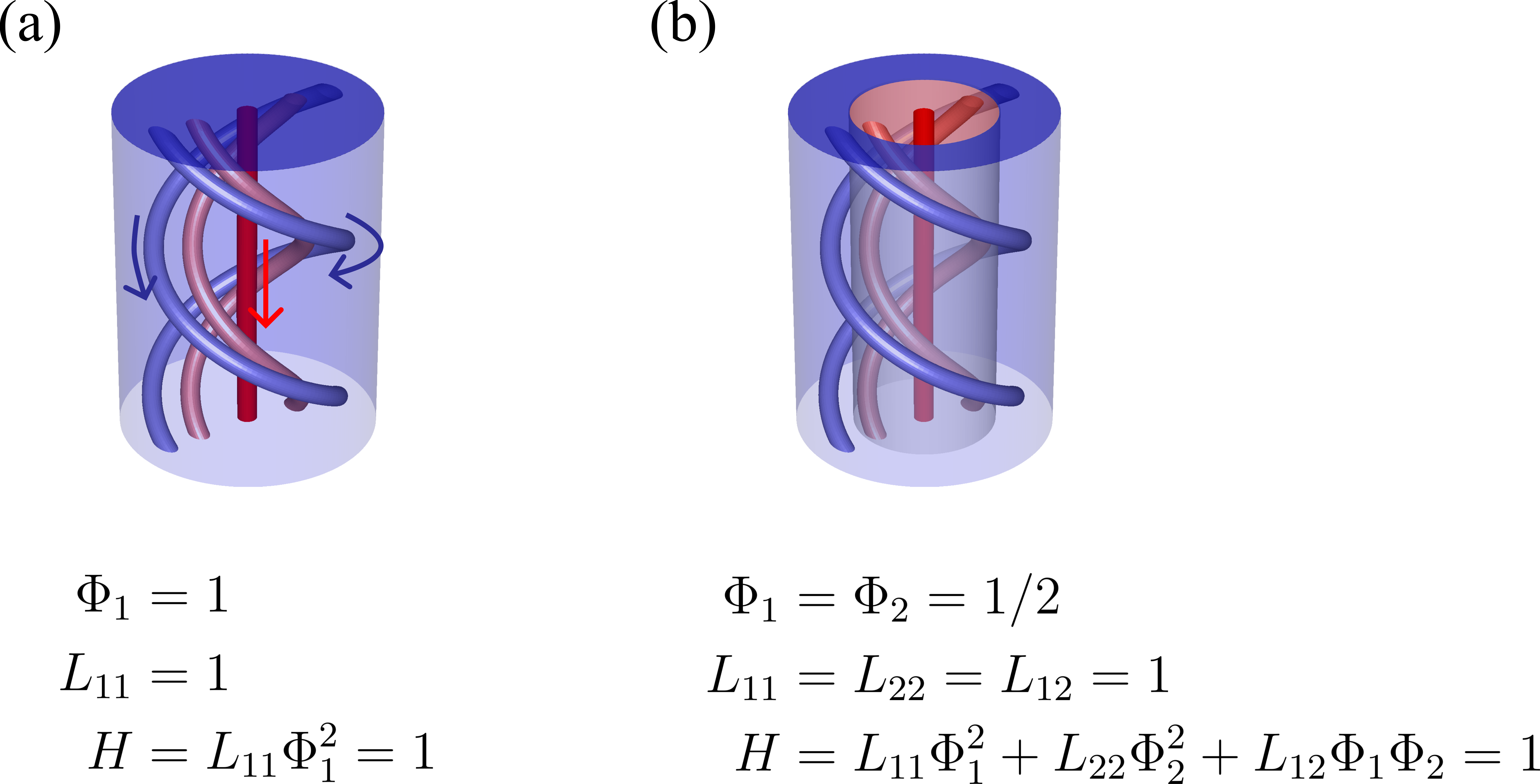}
	\caption{Skyrmion tube in a ferromagnetic background ($m_z^\mathrm{b}=1$) with self-linking, depicted for two different flux tube decompositions:
(a) single flux tube, (b) two concentric flux tubes shown as blue and red surfaces. The blue and red thin tubes depict a selection of the field lines of $\mathbf{F}$, with arrows showing the direction of $\mathbf{F}$.
Independent of the flux tube decomposition, this yields a Hopf index of $H=1$.}
	\label{fig:figsk}
\end{figure}

For a Skyrmion with self-linking, shown in Fig.~\ref{fig:figsk}(a), the Hopf index using Eq.~\eqref{eq:Hopf_decomposed} is given by $H=L_{11}\Phi_1=1$, if we assign all flux to a single flux tube. As an alternative, we can choose a division into two concentric tubes as shown in Fig.~\ref{fig:figsk}(b). The Hopf index is then given by $H=1/4+1/4+1/2=1$.

\label{sm:fluxdecomposition_sk}

\subsection{Hopf index for $N$ number of sub-tubes constituting one Hopfion}
\label{sm:fluxdecomposition}

\begin{figure}
 \includegraphics[width=0.9\columnwidth,left]{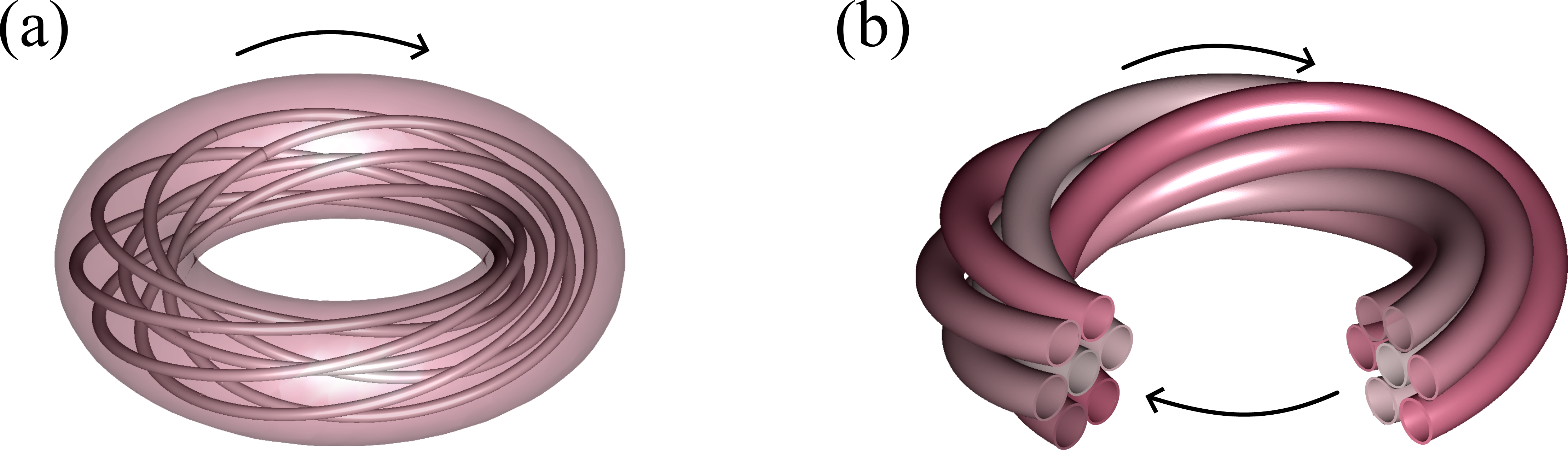}
	\caption{(a) Showing the field lines of emergent magnetic field $\mathbf{F}$ within a single Hopfion, and (b) its division into an arbitrary number of linked magnetic flux tubes. The arrow indicates the direction of $\mathbf{F}$. 
	}
	\label{fig:hopffilamaents}
\end{figure}

Considering a Hopfion as an illustrative example, we show that we can divide a flux tube with flux $\Phi$ into smaller flux filaments. the Hopfion can be considered as a flux tube of emergent magnetic field $\mathbf{F}$ with total flux $\Phi$ and a fixed linking number $L$. The Hopf index shall be, $H=L\Phi^2$. Consider a Hopfion with $L=1$, decomposed into an arbitrary $N$ number of filaments, each with flux $\Phi/N$, as shown schematically in Fig.~\ref{fig:hopffilamaents}, and indexed by the numbers $1$ to $N$. The self-linking as well as the inter-linking numbers shall be $L_{11}=L_{22}=...=L_{NN}=L_{12}=...=L_{N-1,N}=1$. The Hopf index for the collection of these $N$ filaments is
\begin{align}
    H&=N \left(\frac{\Phi}{N} \right)^2 +2\binom{N}{2} (1)\left(\frac{\Phi}{N} \right)^2 \\ \notag
    &=\Phi^2,
\end{align}
where $\binom{N}{2}$ is the binomial coefficient counting the number of possible pairs of filaments. This result
gives the expected value $H=1$ for $\Phi=1$, showing that Eq.~\eqref{eq:Hopf_decomposed} correctly calculates the Hopf index for a Hopfion decomposed into $N$ number of self-linked and inter-linked flux filaments.

\subsection{Topological decomposition for a Heliknoton}
\begin{figure}
\includegraphics[width=0.85\columnwidth]{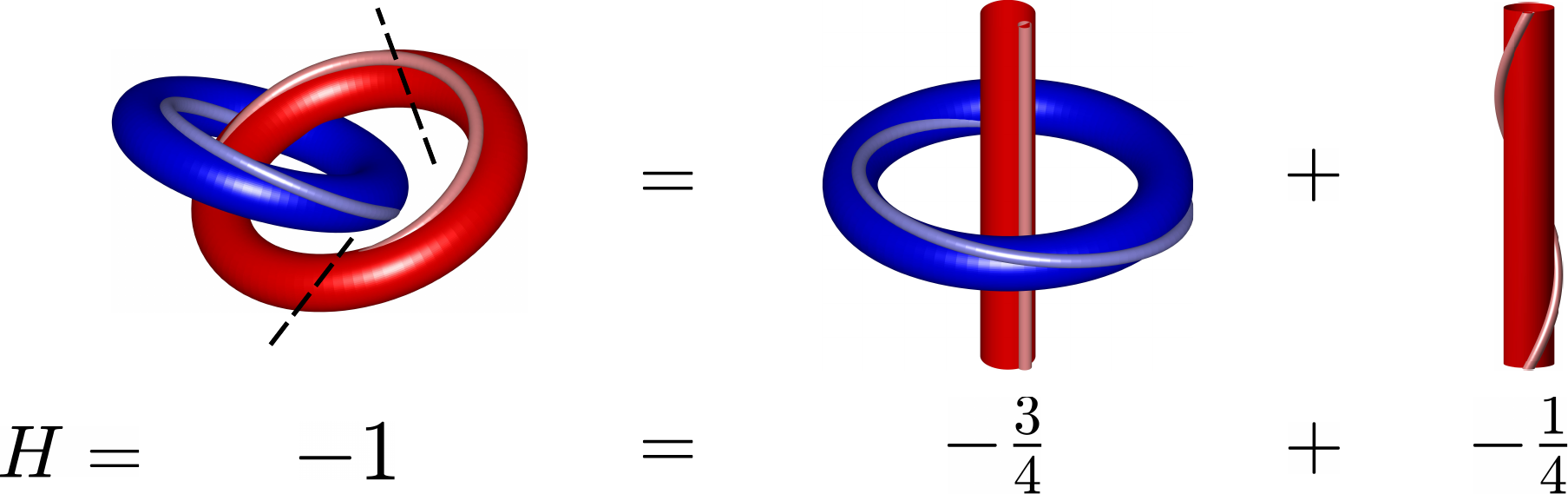}
	\caption{The emergent magnetic field $\mathbf{F}$ comprising the Heliknoton can be cut along the dashed lines to be decomposed into the sum of two constituents: the Twiston and an antivortex meron with a self-linking. There are periodic boundary conditions along the $z$ direction in all cases. The blue and red tubes represent flux tubes of $\mathbf{F}$ with $m_z>0$ and $m_z<0$ in their interior respectively. The light blue(red) line represents one field line of $\mathbf{F}$.}
	\label{fig:topdecompose}
\end{figure}

Magnetic textures can be decomposed into their constituent topological textures. As an example, consider the Heliknoton (Fig.~\ref{fig:spiralbackground}(c)) that can be decomposed into the sum of a screw dislocation with $\nu=-1$  (Fig.~\ref{fig:spiralbackground}(e)) and a screw dislocation with $\nu=1$ (Fig.~\ref{fig:spiralbackground}(f)). The Hopf index sums to $-1$ at any value of $m_z^\mathrm{b}$, demonstrating the additivity of the $H$ when two simply connected volumes are merged while deforming $\mathbf{F}$ near the surface, to preserve the continuity of $\mathbf{F}$. An illustration of this concept is shown in Fig.~\ref{fig:topdecompose} where the $\mathbf{F}$ constituting the Heliknoton is shown to be split into two components.

\section{Additional numerics}
\label{supp:numerics}

\subsection{Numerically relaxed magnetic textures}

\begin{figure*}
	\includegraphics[width=1\textwidth]{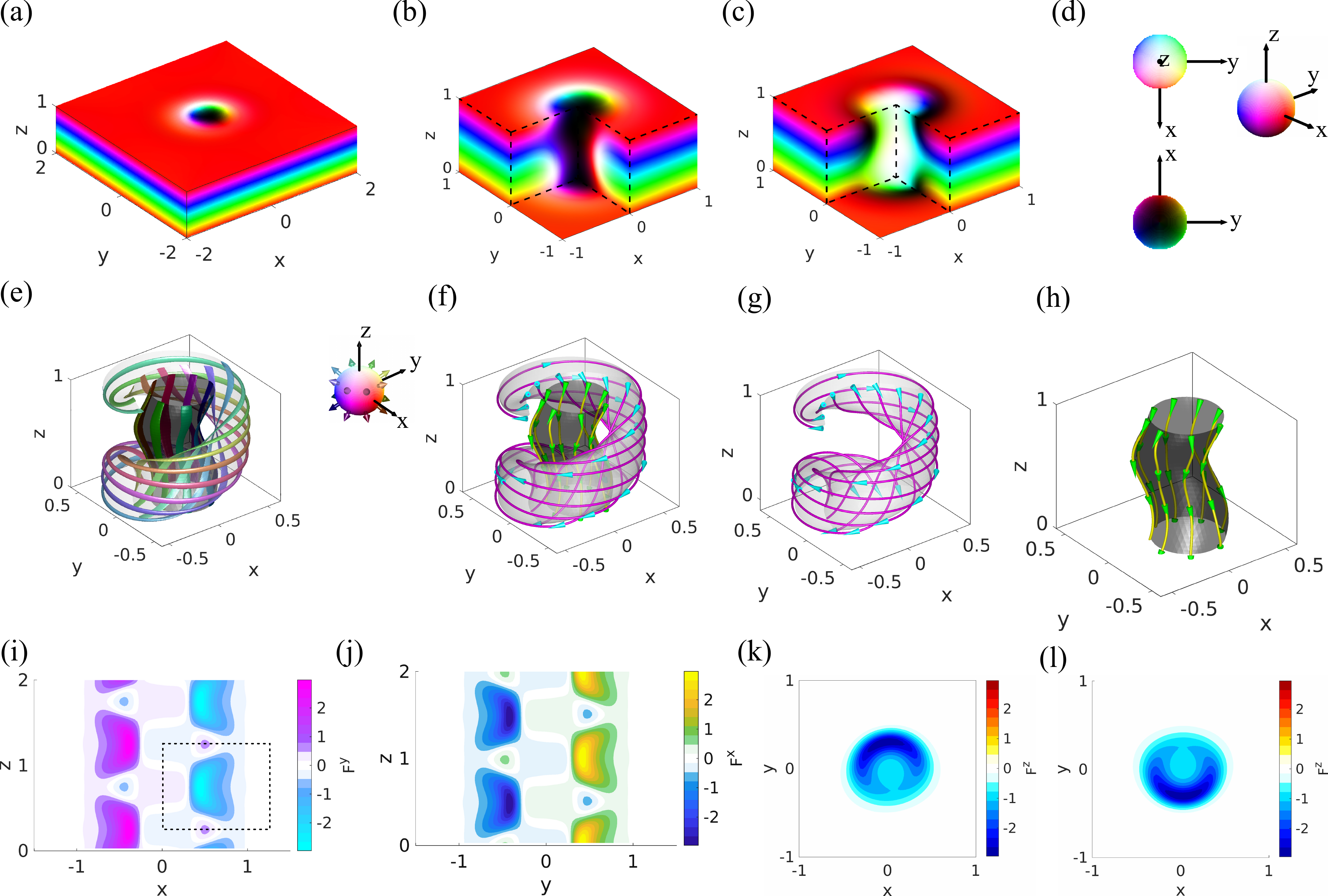}
	\caption{Numerically relaxed magnetisation configuration of the Skyrmion string aligned along the $z$-axis in a helical spiral background with spiral wavevector also along the $z$ direction. (a) displays a box of height $L_{\mathrm{D}}$, centred on the Skyrmion  string at $(x,y)=(0,0)$. The HSL colourmap shown in (d) is used to represent the magnetisation direction $\mathbf{m}$ in panels (a) to (c). In (a) and (b) the $xy$-plane features a vortex (black) and an antivortex tube (white), with cores polarized along $-z$ and $+z$ respectively.  (b) and (c) show cuts through the centre of the box, for Skyrmions with $(N_{\mathrm{sk}},H)=(-1,-1)$ and $(N_{\mathrm{sk}},H)=(1,-1)$, respectively, that are related simply by a transformation $\mathbf{m}\rightarrow-\mathbf{m}$, $z\rightarrow z+0.5$. (e) shows selected preimages of magnetisation for $m_z=\pm0.5$, overlaid on the isosurfaces of $m_z=\pm0.5$. (f) depicts the field lines of the emergent magnetic field $\mathbf{F}$, overlaid on the isosurfaces of $m_z=\pm0.5$. The arrows depict the direction of $\mathbf{F}$. (g) and (h) separately display the data from (f) for $m_z=+0.5$ and $m_z=-0.5$, respectively. (i) and (j) present plots of $F_x$ and $F_y$, respectively, with each half-plane (with either $x>0$, $x<0$, $y>0$ or $y<0$) pierced by a total flux $\Phi=1/2$, per period along $z$. The dashed line in (i) and (j) encloses a region with $\Phi=1/2$. (k) and (l) show $F_z$ in the planes $z=0$ and $z=0.5$, respectively. All lengths are measured in units of the spiral period.	}
	\label{fig:sk}
 \end{figure*}

\begin{figure*}
 \includegraphics[width=1\textwidth]{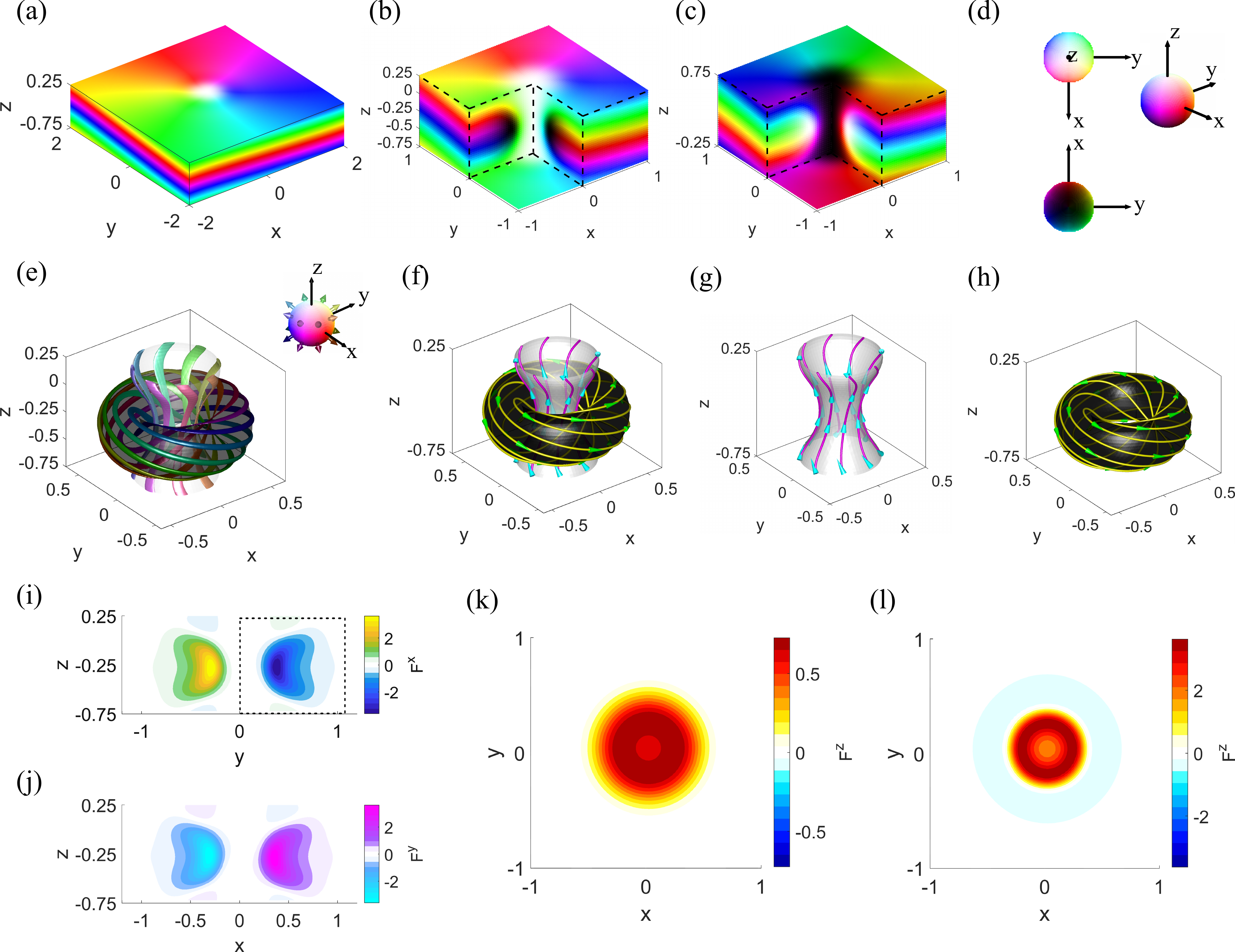}
	\caption{Numerically relaxed magnetisation configuration of the Twiston with $\nu=+1$ described by the asymptotic behavior $\phi=\nu\chi+qz$ for the far field. (a) displays a box of height $L_{\mathrm{D}}$. The HSL colourmap shown in (d) is used to represent the magnetisation direction $\mathbf{m}$ in panels (a) to (c).   (b) and (c) show cuts through the centre of the box, for textures with $(N_{\mathrm{sk}},H)=(-p/2,3/4)$ where $p=+1$ and $p=-1$ respectively, that are related simply by a transformation $\mathbf{m}\rightarrow-\mathbf{m}$. (e) shows selected preimages of magnetisation for $m_z=\pm0.5$, overlaid on the isosurfaces of $m_z=\pm0.5$. (f) depicts the field lines of the emergent magnetic field $\mathbf{F}$, overlaid on the isosurfaces of $m_z=\pm0.5$. (g) and (h) separately display the data from (f) for $m_z=+0.5$ and $m_z=-0.5$, respectively. (i) and (j) present plots of $F_x$ and $F_y$, respectively, with each half-plane (with either $x>0$, $x<0$, $y>0$ or $y<0$) pierced by a total flux $\Phi=1/2$, per period along $z$. The dashed line in (i) encloses a region with $\Phi=1/2$. (k) and (l) show $F_z$ in the planes $z=0.25$ and $z=-0.25$, respectively.	 All lengths are measured in units of the spiral period.}
	\label{fig:sdhopf}
\end{figure*}

We show details of numerically relaxed textures, the Skyrmion in a spiral background (Fig.~\ref{fig:sk}) and the Twiston (Fig.~\ref{fig:sdhopf}).
The vector field $\mathbf{F}$ is plotted for cross-sections in the $yz$-, $xz$- and $xy$-planes in panels (i) to (l) of Figs.~\ref{fig:sk} and \ref{fig:sdhopf}. Along the $x$ and $y$ directions, $\mathbf{F}\rightarrow\mathbf{0}$ as $x\rightarrow\infty$ or $y\rightarrow\infty$ for all textures considered.

\hfill \break

\subsection{Choice of gauge potential}

The Hopf index, calculated  using the Whitehead equation  $H=-\int \mathrm{d}^3r \, \mathbf{F}\cdot\mathbf{A}$  for a finite integration volume $V$, is gauge invariant when $\mathbf{F}\cdot\hat{n}$ vanishes at the surfaces of $V$ (where $\hat{n}$ is the surface normal) \cite{Knapman2024a}. However, if $\mathbf{F}$ does not satisfy these boundary conditions --- for example, when the field lines of $\mathbf{F}$ align along the $z$-axis with periodic boundary conditions along $z$ --- then $H$ is a topological invariant only when it is measured relative to a reference state~\cite{Berger1984,Prior2014}. This condition restricts the acceptable choices for the gauge $\mathbf{A}$.  


As a natural reference state, we consider a translationally invariant Skyrmion tube which must have the trivial Hopf index of $0$ (see Fig.~\ref{fig:examplesfm}(a)) using the chosen gauge. In addition, when $\mathbf{F}$ is periodic, $\mathbf{A}$ must satisfy the same periodic boundary conditions. A gauge that satisfies these conditions is $A_y=0$, and inverting the equation $\mathbf{\nabla}\times\mathbf{A}=\mathbf{F}$ gives,
\begin{equation}
\begin{aligned}
	A_x(x, y, z) &=- \int_{-L_{y/2}}^y \mathrm{d}y' F_z(x, y', z), \\
	A_y(x, y, z) &= 0, \\
	A_z(x, y, z) &= \int_{-L_{y/2}}^y \mathrm{d}y' F_x(x, y', z).
\end{aligned}
\label{eq:VectorPotential}
\end{equation}
 This gauge is used for our numerical calculations of $H$, consistent with Ref.~\cite{Knapman2024a}.  $L_x$($L_y$) are dimensions of the integration volume along the $x$($y$)-direction. $L_x$ and $L_y$ must be large enough so that $\mathbf{F}\approx\mathbf{0}$ at $x=\pm L_x/2$ and $y=\pm L_y/2$. For the results in Fig.~\ref{fig:Numerics_spiralbackground}, sufficiently accurate results for $H$ are obtained by using an integration volume with $L_x=L_y=5L_D$ where $L_D$ is the size of the integration volume along the $z$ direction.  $\mathbf{F}$ is calculated using the average solid angle between neighboring lattice sites \cite{Knapman2024a}.

\begin{thebibliography}{75}%
	\makeatletter
	\providecommand \@ifxundefined [1]{%
		\@ifx{#1\undefined}
	}%
	\providecommand \@ifnum [1]{%
		\ifnum #1\expandafter \@firstoftwo
		\else \expandafter \@secondoftwo
		\fi
	}%
	\providecommand \@ifx [1]{%
		\ifx #1\expandafter \@firstoftwo
		\else \expandafter \@secondoftwo
		\fi
	}%
	\providecommand \natexlab [1]{#1}%
	\providecommand \enquote  [1]{``#1''}%
	\providecommand \bibnamefont  [1]{#1}%
	\providecommand \bibfnamefont [1]{#1}%
	\providecommand \citenamefont [1]{#1}%
	\providecommand \href@noop [0]{\@secondoftwo}%
	\providecommand \href [0]{\begingroup \@sanitize@url \@href}%
	\providecommand \@href[1]{\@@startlink{#1}\@@href}%
	\providecommand \@@href[1]{\endgroup#1\@@endlink}%
	\providecommand \@sanitize@url [0]{\catcode `\\12\catcode `\$12\catcode
		`\&12\catcode `\#12\catcode `\^12\catcode `\_12\catcode `\%12\relax}%
	\providecommand \@@startlink[1]{}%
	\providecommand \@@endlink[0]{}%
	\providecommand \url  [0]{\begingroup\@sanitize@url \@url }%
	\providecommand \@url [1]{\endgroup\@href {#1}{\urlprefix }}%
	\providecommand \urlprefix  [0]{URL }%
	\providecommand \Eprint [0]{\href }%
	\providecommand \doibase [0]{https://doi.org/}%
	\providecommand \selectlanguage [0]{\@gobble}%
	\providecommand \bibinfo  [0]{\@secondoftwo}%
	\providecommand \bibfield  [0]{\@secondoftwo}%
	\providecommand \translation [1]{[#1]}%
	\providecommand \BibitemOpen [0]{}%
	\providecommand \bibitemStop [0]{}%
	\providecommand \bibitemNoStop [0]{.\EOS\space}%
	\providecommand \EOS [0]{\spacefactor3000\relax}%
	\providecommand \BibitemShut  [1]{\csname bibitem#1\endcsname}%
	\let\auto@bib@innerbib\@empty
	\bibitem [{\citenamefont {Ricca}\ and\ \citenamefont
		{Liu}(2024)}]{KnottedFields2024}%
	\BibitemOpen
	\bibinfo {editor} {\bibfnamefont {R.~L.}\ \bibnamefont {Ricca}}\ and\
	\bibinfo {editor} {\bibfnamefont {X.}~\bibnamefont {Liu}},\ eds.,\ \href
	{https://doi.org/10.1007/978-3-031-57985-1} {\emph {\bibinfo {title} {Knotted
				Fields}}},\ \bibinfo {series} {Lecture Notes in Mathematics}, Vol.\ \bibinfo
	{volume} {2344}\ (\bibinfo  {publisher} {Springer},\ \bibinfo {address}
	{Cham},\ \bibinfo {year} {2024})\BibitemShut {NoStop}%
	\bibitem [{\citenamefont {Kleckner}\ and\ \citenamefont
		{Irvine}(2013)}]{Kleckner2013}%
	\BibitemOpen
	\bibfield  {author} {\bibinfo {author} {\bibfnamefont {D.}~\bibnamefont
			{Kleckner}}\ and\ \bibinfo {author} {\bibfnamefont {W.~T.~M.}\ \bibnamefont
			{Irvine}},\ }\href {https://doi.org/10.1038/nphys2560} {\bibfield  {journal}
		{\bibinfo  {journal} {Nat. Phys.}\ }\textbf {\bibinfo {volume} {9}},\
		\bibinfo {pages} {253} (\bibinfo {year} {2013})}\BibitemShut {NoStop}%
	\bibitem [{\citenamefont {Neophytou}\ \emph {et~al.}(2022)\citenamefont
		{Neophytou}, \citenamefont {Chakrabarti},\ and\ \citenamefont
		{Sciortino}}]{Neophytou2022}%
	\BibitemOpen
	\bibfield  {author} {\bibinfo {author} {\bibfnamefont {A.}~\bibnamefont
			{Neophytou}}, \bibinfo {author} {\bibfnamefont {D.}~\bibnamefont
			{Chakrabarti}},\ and\ \bibinfo {author} {\bibfnamefont {F.}~\bibnamefont
			{Sciortino}},\ }\href {https://doi.org/10.1038/s41567-022-01698-6} {\bibfield
		{journal} {\bibinfo  {journal} {Nature Physics}\ }\textbf {\bibinfo {volume}
			{18}},\ \bibinfo {pages} {1248} (\bibinfo {year} {2022})}\BibitemShut
	{NoStop}%
	\bibitem [{\citenamefont {Ollikainen}\ \emph {et~al.}(2019)\citenamefont
		{Ollikainen}, \citenamefont {Blinova}, \citenamefont {M\"ott\"onen},\ and\
		\citenamefont {Hall}}]{Ollikainen2019}%
	\BibitemOpen
	\bibfield  {author} {\bibinfo {author} {\bibfnamefont {T.}~\bibnamefont
			{Ollikainen}}, \bibinfo {author} {\bibfnamefont {A.}~\bibnamefont {Blinova}},
		\bibinfo {author} {\bibfnamefont {M.}~\bibnamefont {M\"ott\"onen}},\ and\
		\bibinfo {author} {\bibfnamefont {D.~S.}\ \bibnamefont {Hall}},\ }\href
	{https://doi.org/10.1103/PhysRevLett.123.163003} {\bibfield  {journal}
		{\bibinfo  {journal} {Phys. Rev. Lett.}\ }\textbf {\bibinfo {volume} {123}},\
		\bibinfo {pages} {163003} (\bibinfo {year} {2019})}\BibitemShut {NoStop}%
	\bibitem [{\citenamefont {Kedia}\ \emph {et~al.}(2013)\citenamefont {Kedia},
		\citenamefont {Bialynicki-Birula}, \citenamefont {Peralta-Salas},\ and\
		\citenamefont {Irvine}}]{Kedia2013}%
	\BibitemOpen
	\bibfield  {author} {\bibinfo {author} {\bibfnamefont {H.}~\bibnamefont
			{Kedia}}, \bibinfo {author} {\bibfnamefont {I.}~\bibnamefont
			{Bialynicki-Birula}}, \bibinfo {author} {\bibfnamefont {D.}~\bibnamefont
			{Peralta-Salas}},\ and\ \bibinfo {author} {\bibfnamefont {W.~T.~M.}\
			\bibnamefont {Irvine}},\ }\href
	{https://doi.org/10.1103/PhysRevLett.111.150404} {\bibfield  {journal}
		{\bibinfo  {journal} {Phys. Rev. Lett.}\ }\textbf {\bibinfo {volume} {111}},\
		\bibinfo {pages} {150404} (\bibinfo {year} {2013})}\BibitemShut {NoStop}%
	\bibitem [{\citenamefont {Han}\ \emph {et~al.}(2010)\citenamefont {Han},
		\citenamefont {Pal}, \citenamefont {Liu},\ and\ \citenamefont
		{Yan}}]{Han2010}%
	\BibitemOpen
	\bibfield  {author} {\bibinfo {author} {\bibfnamefont {D.}~\bibnamefont
			{Han}}, \bibinfo {author} {\bibfnamefont {S.}~\bibnamefont {Pal}}, \bibinfo
		{author} {\bibfnamefont {Y.}~\bibnamefont {Liu}},\ and\ \bibinfo {author}
		{\bibfnamefont {H.}~\bibnamefont {Yan}},\ }\href
	{https://doi.org/10.1038/nnano.2010.193} {\bibfield  {journal} {\bibinfo
			{journal} {Nat. Nanotechnol.}\ }\textbf {\bibinfo {volume} {5}},\ \bibinfo
		{pages} {712} (\bibinfo {year} {2010})}\BibitemShut {NoStop}%
	\bibitem [{\citenamefont {Kleckner}\ \emph {et~al.}(2016)\citenamefont
		{Kleckner}, \citenamefont {Kauffman},\ and\ \citenamefont
		{Irvine}}]{Kleckner2016}%
	\BibitemOpen
	\bibfield  {author} {\bibinfo {author} {\bibfnamefont {D.}~\bibnamefont
			{Kleckner}}, \bibinfo {author} {\bibfnamefont {L.~H.}\ \bibnamefont
			{Kauffman}},\ and\ \bibinfo {author} {\bibfnamefont {W.~T.~M.}\ \bibnamefont
			{Irvine}},\ }\href {https://doi.org/10.1038/nphys3679} {\bibfield  {journal}
		{\bibinfo  {journal} {Nat. Phys.}\ }\textbf {\bibinfo {volume} {12}},\
		\bibinfo {pages} {650} (\bibinfo {year} {2016})}\BibitemShut {NoStop}%
	\bibitem [{\citenamefont {Tkalec}\ \emph {et~al.}(2011)\citenamefont {Tkalec},
		\citenamefont {Ravnik}, \citenamefont {{\v{C}}opar}, \citenamefont
		{{\v{Z}}umer},\ and\ \citenamefont {Mu{\v{s}}evi{\v{c}}}}]{Tkalec2011}%
	\BibitemOpen
	\bibfield  {author} {\bibinfo {author} {\bibfnamefont {U.}~\bibnamefont
			{Tkalec}}, \bibinfo {author} {\bibfnamefont {M.}~\bibnamefont {Ravnik}},
		\bibinfo {author} {\bibfnamefont {S.}~\bibnamefont {{\v{C}}opar}}, \bibinfo
		{author} {\bibfnamefont {S.}~\bibnamefont {{\v{Z}}umer}},\ and\ \bibinfo
		{author} {\bibfnamefont {I.}~\bibnamefont {Mu{\v{s}}evi{\v{c}}}},\ }\href
	{https://doi.org/10.1126/science.1205705} {\bibfield  {journal} {\bibinfo
			{journal} {Science}\ }\textbf {\bibinfo {volume} {333}},\ \bibinfo {pages}
		{62} (\bibinfo {year} {2011})}\BibitemShut {NoStop}%
	\bibitem [{\citenamefont {Binysh}\ \emph {et~al.}(2020)\citenamefont {Binysh},
		\citenamefont {Pollard},\ and\ \citenamefont {Alexander}}]{Binysh2020}%
	\BibitemOpen
	\bibfield  {author} {\bibinfo {author} {\bibfnamefont {J.}~\bibnamefont
			{Binysh}}, \bibinfo {author} {\bibfnamefont {J.}~\bibnamefont {Pollard}},\
		and\ \bibinfo {author} {\bibfnamefont {G.~P.}\ \bibnamefont {Alexander}},\
	}\href {https://doi.org/10.1103/PhysRevLett.125.047801} {\bibfield  {journal}
		{\bibinfo  {journal} {Phys. Rev. Lett.}\ }\textbf {\bibinfo {volume} {125}},\
		\bibinfo {pages} {047801} (\bibinfo {year} {2020})}\BibitemShut {NoStop}%
	\bibitem [{\citenamefont {Smalyukh}(2020)}]{Smalyukh_2020}%
	\BibitemOpen
	\bibfield  {author} {\bibinfo {author} {\bibfnamefont {I.~I.}\ \bibnamefont
			{Smalyukh}},\ }\href {https://doi.org/10.1088/1361-6633/abaa39} {\bibfield
		{journal} {\bibinfo  {journal} {Rep. Prog. Phys.}\ }\textbf {\bibinfo
			{volume} {83}},\ \bibinfo {pages} {106601} (\bibinfo {year}
		{2020})}\BibitemShut {NoStop}%
	\bibitem [{\citenamefont {Wu}\ and\ \citenamefont {Smalyukh}(2022)}]{Wu2022}%
	\BibitemOpen
	\bibfield  {author} {\bibinfo {author} {\bibfnamefont {J.-S.}\ \bibnamefont
			{Wu}}\ and\ \bibinfo {author} {\bibfnamefont {I.~I.}\ \bibnamefont
			{Smalyukh}},\ }\href {https://doi.org/10.1080/21680396.2022.2040058}
	{\bibfield  {journal} {\bibinfo  {journal} {Liq. Cryst. Rev.}\ }\textbf
		{\bibinfo {volume} {10}},\ \bibinfo {pages} {34} (\bibinfo {year}
		{2022})}\BibitemShut {NoStop}%
	\bibitem [{\citenamefont {Sugic}\ \emph {et~al.}(2021)\citenamefont {Sugic},
		\citenamefont {Droop}, \citenamefont {Otte}, \citenamefont {Ehrmanntraut},
		\citenamefont {Nori}, \citenamefont {Ruostekoski}, \citenamefont {Denz},\
		and\ \citenamefont {Dennis}}]{Sugic2021}%
	\BibitemOpen
	\bibfield  {author} {\bibinfo {author} {\bibfnamefont {D.}~\bibnamefont
			{Sugic}}, \bibinfo {author} {\bibfnamefont {R.}~\bibnamefont {Droop}},
		\bibinfo {author} {\bibfnamefont {E.}~\bibnamefont {Otte}}, \bibinfo {author}
		{\bibfnamefont {D.}~\bibnamefont {Ehrmanntraut}}, \bibinfo {author}
		{\bibfnamefont {F.}~\bibnamefont {Nori}}, \bibinfo {author} {\bibfnamefont
			{J.}~\bibnamefont {Ruostekoski}}, \bibinfo {author} {\bibfnamefont
			{C.}~\bibnamefont {Denz}},\ and\ \bibinfo {author} {\bibfnamefont {M.~R.}\
			\bibnamefont {Dennis}},\ }\href {https://doi.org/10.1038/s41467-021-26171-5}
	{\bibfield  {journal} {\bibinfo  {journal} {Nat. Commun.}\ }\textbf {\bibinfo
			{volume} {12}},\ \bibinfo {pages} {6785} (\bibinfo {year}
		{2021})}\BibitemShut {NoStop}%
	\bibitem [{\citenamefont {Kong}\ \emph {et~al.}(2022)\citenamefont {Kong},
		\citenamefont {Zhang}, \citenamefont {Li}, \citenamefont {Wang},
		\citenamefont {Meng}, \citenamefont {Jiang}, \citenamefont {Zhang},
		\citenamefont {Liu}, \citenamefont {Yang}, \citenamefont {Sun}, \citenamefont
		{Zhang}, \citenamefont {Shen}, \citenamefont {Li}, \citenamefont {Fang},
		\citenamefont {Chen},\ and\ \citenamefont {Wang}}]{Kong2022}%
	\BibitemOpen
	\bibfield  {author} {\bibinfo {author} {\bibfnamefont {L.}~\bibnamefont
			{Kong}}, \bibinfo {author} {\bibfnamefont {W.}~\bibnamefont {Zhang}},
		\bibinfo {author} {\bibfnamefont {P.}~\bibnamefont {Li}}, \bibinfo {author}
		{\bibfnamefont {Z.}~\bibnamefont {Wang}}, \bibinfo {author} {\bibfnamefont
			{C.}~\bibnamefont {Meng}}, \bibinfo {author} {\bibfnamefont {H.}~\bibnamefont
			{Jiang}}, \bibinfo {author} {\bibfnamefont {Y.}~\bibnamefont {Zhang}},
		\bibinfo {author} {\bibfnamefont {S.}~\bibnamefont {Liu}}, \bibinfo {author}
		{\bibfnamefont {S.}~\bibnamefont {Yang}}, \bibinfo {author} {\bibfnamefont
			{J.}~\bibnamefont {Sun}}, \bibinfo {author} {\bibfnamefont {S.}~\bibnamefont
			{Zhang}}, \bibinfo {author} {\bibfnamefont {Y.}~\bibnamefont {Shen}},
		\bibinfo {author} {\bibfnamefont {X.}~\bibnamefont {Li}}, \bibinfo {author}
		{\bibfnamefont {X.}~\bibnamefont {Fang}}, \bibinfo {author} {\bibfnamefont
			{X.}~\bibnamefont {Chen}},\ and\ \bibinfo {author} {\bibfnamefont
			{J.}~\bibnamefont {Wang}},\ }\href
	{https://doi.org/10.1038/s41467-022-30381-w} {\bibfield  {journal} {\bibinfo
			{journal} {Nat. Commun.}\ }\textbf {\bibinfo {volume} {13}},\ \bibinfo
		{pages} {2705} (\bibinfo {year} {2022})}\BibitemShut {NoStop}%
	\bibitem [{\citenamefont {Faddeev}\ and\ \citenamefont
		{Niemi}(1997)}]{Faddeev1997}%
	\BibitemOpen
	\bibfield  {author} {\bibinfo {author} {\bibfnamefont {L.}~\bibnamefont
			{Faddeev}}\ and\ \bibinfo {author} {\bibfnamefont {A.~J.}\ \bibnamefont
			{Niemi}},\ }\href {https://doi.org/10.1038/387058a0} {\bibfield  {journal}
		{\bibinfo  {journal} {Nature}\ }\textbf {\bibinfo {volume} {387}},\ \bibinfo
		{pages} {58} (\bibinfo {year} {1997})}\BibitemShut {NoStop}%
	\bibitem [{\citenamefont {Sutcliffe}(2017)}]{Sutcliffe2017}%
	\BibitemOpen
	\bibfield  {author} {\bibinfo {author} {\bibfnamefont {P.}~\bibnamefont
			{Sutcliffe}},\ }\href {https://doi.org/10.1103/PhysRevLett.118.247203}
	{\bibfield  {journal} {\bibinfo  {journal} {Phys. Rev. Lett.}\ }\textbf
		{\bibinfo {volume} {118}},\ \bibinfo {pages} {247203} (\bibinfo {year}
		{2017})}\BibitemShut {NoStop}%
	\bibitem [{\citenamefont {Rybakov}\ \emph {et~al.}(2022)\citenamefont
		{Rybakov}, \citenamefont {Kiselev}, \citenamefont {Borisov}, \citenamefont
		{D{\"o}ring}, \citenamefont {Melcher},\ and\ \citenamefont
		{Bl{\"u}gel}}]{Rybakov2022}%
	\BibitemOpen
	\bibfield  {author} {\bibinfo {author} {\bibfnamefont {F.~N.}\ \bibnamefont
			{Rybakov}}, \bibinfo {author} {\bibfnamefont {N.~S.}\ \bibnamefont
			{Kiselev}}, \bibinfo {author} {\bibfnamefont {A.~B.}\ \bibnamefont
			{Borisov}}, \bibinfo {author} {\bibfnamefont {L.}~\bibnamefont {D{\"o}ring}},
		\bibinfo {author} {\bibfnamefont {C.}~\bibnamefont {Melcher}},\ and\ \bibinfo
		{author} {\bibfnamefont {S.}~\bibnamefont {Bl{\"u}gel}},\ }\href
	{https://doi.org/10.1063/5.0099942} {\bibfield  {journal} {\bibinfo
			{journal} {APL Mater.}\ }\textbf {\bibinfo {volume} {10}},\ \bibinfo {pages}
		{111113} (\bibinfo {year} {2022})}\BibitemShut {NoStop}%
	\bibitem [{\citenamefont {Kent}\ \emph {et~al.}(2021)\citenamefont {Kent},
		\citenamefont {Reynolds}, \citenamefont {Raftrey}, \citenamefont {Campbell},
		\citenamefont {Virasawmy}, \citenamefont {Dhuey}, \citenamefont {Chopdekar},
		\citenamefont {Hierro-Rodriguez}, \citenamefont {Sorrentino}, \citenamefont
		{Pereiro}, \citenamefont {Ferrer}, \citenamefont {Hellman}, \citenamefont
		{Sutcliffe},\ and\ \citenamefont {Fischer}}]{Kent2021}%
	\BibitemOpen
	\bibfield  {author} {\bibinfo {author} {\bibfnamefont {N.}~\bibnamefont
			{Kent}}, \bibinfo {author} {\bibfnamefont {N.}~\bibnamefont {Reynolds}},
		\bibinfo {author} {\bibfnamefont {D.}~\bibnamefont {Raftrey}}, \bibinfo
		{author} {\bibfnamefont {I.~T.~G.}\ \bibnamefont {Campbell}}, \bibinfo
		{author} {\bibfnamefont {S.}~\bibnamefont {Virasawmy}}, \bibinfo {author}
		{\bibfnamefont {S.}~\bibnamefont {Dhuey}}, \bibinfo {author} {\bibfnamefont
			{R.~V.}\ \bibnamefont {Chopdekar}}, \bibinfo {author} {\bibfnamefont
			{A.}~\bibnamefont {Hierro-Rodriguez}}, \bibinfo {author} {\bibfnamefont
			{A.}~\bibnamefont {Sorrentino}}, \bibinfo {author} {\bibfnamefont
			{E.}~\bibnamefont {Pereiro}}, \bibinfo {author} {\bibfnamefont
			{S.}~\bibnamefont {Ferrer}}, \bibinfo {author} {\bibfnamefont
			{F.}~\bibnamefont {Hellman}}, \bibinfo {author} {\bibfnamefont
			{P.}~\bibnamefont {Sutcliffe}},\ and\ \bibinfo {author} {\bibfnamefont
			{P.}~\bibnamefont {Fischer}},\ }\href
	{https://doi.org/10.1038/s41467-021-21846-5} {\bibfield  {journal} {\bibinfo
			{journal} {Nature Communications}\ }\textbf {\bibinfo {volume} {12}},\
		\bibinfo {pages} {1562} (\bibinfo {year} {2021})}\BibitemShut {NoStop}%
	\bibitem [{\citenamefont {Tai}\ and\ \citenamefont {Smalyukh}(2018)}]{Tai18}%
	\BibitemOpen
	\bibfield  {author} {\bibinfo {author} {\bibfnamefont {J.-S.~B.}\
			\bibnamefont {Tai}}\ and\ \bibinfo {author} {\bibfnamefont {I.~I.}\
			\bibnamefont {Smalyukh}},\ }\href
	{https://doi.org/10.1103/physrevlett.121.187201} {\bibfield  {journal}
		{\bibinfo  {journal} {Physical Review Letters}\ }\textbf {\bibinfo {volume}
			{121}},\ \bibinfo {pages} {187201} (\bibinfo {year} {2018})}\BibitemShut
	{NoStop}%
	\bibitem [{\citenamefont {Sutcliffe}(2018)}]{Sutcliffe18}%
	\BibitemOpen
	\bibfield  {author} {\bibinfo {author} {\bibfnamefont {P.}~\bibnamefont
			{Sutcliffe}},\ }\href {https://doi.org/10.1088/1751-8121/aad521} {\bibfield
		{journal} {\bibinfo  {journal} {J. Phys. A: Math. Theor.}\ }\textbf {\bibinfo
			{volume} {51}},\ \bibinfo {pages} {375401} (\bibinfo {year}
		{2018})}\BibitemShut {NoStop}%
	\bibitem [{\citenamefont {Voinescu}\ \emph {et~al.}(2020)\citenamefont
		{Voinescu}, \citenamefont {Tai},\ and\ \citenamefont
		{Smalyukh}}]{Voinescu2020}%
	\BibitemOpen
	\bibfield  {author} {\bibinfo {author} {\bibfnamefont {R.}~\bibnamefont
			{Voinescu}}, \bibinfo {author} {\bibfnamefont {J.-S.~B.}\ \bibnamefont
			{Tai}},\ and\ \bibinfo {author} {\bibfnamefont {I.~I.}\ \bibnamefont
			{Smalyukh}},\ }\href {https://doi.org/10.1103/PhysRevLett.125.057201}
	{\bibfield  {journal} {\bibinfo  {journal} {Phys. Rev. Lett.}\ }\textbf
		{\bibinfo {volume} {125}},\ \bibinfo {pages} {057201} (\bibinfo {year}
		{2020})}\BibitemShut {NoStop}%
	\bibitem [{\citenamefont {Azhar}\ \emph {et~al.}(2022)\citenamefont {Azhar},
		\citenamefont {Kravchuk},\ and\ \citenamefont {Garst}}]{Azhar2022}%
	\BibitemOpen
	\bibfield  {author} {\bibinfo {author} {\bibfnamefont {M.}~\bibnamefont
			{Azhar}}, \bibinfo {author} {\bibfnamefont {V.~P.}\ \bibnamefont
			{Kravchuk}},\ and\ \bibinfo {author} {\bibfnamefont {M.}~\bibnamefont
			{Garst}},\ }\href {https://doi.org/10.1103/PhysRevLett.128.157204} {\bibfield
		{journal} {\bibinfo  {journal} {Phys. Rev. Lett.}\ }\textbf {\bibinfo
			{volume} {128}},\ \bibinfo {pages} {157204} (\bibinfo {year}
		{2022})}\BibitemShut {NoStop}%
	\bibitem [{\citenamefont {Zheng}\ \emph {et~al.}(2023)\citenamefont {Zheng},
		\citenamefont {Kiselev}, \citenamefont {Rybakov}, \citenamefont {Yang},
		\citenamefont {Shi}, \citenamefont {Bl{\"u}gel},\ and\ \citenamefont
		{Dunin-Borkowski}}]{Zheng2023}%
	\BibitemOpen
	\bibfield  {author} {\bibinfo {author} {\bibfnamefont {F.}~\bibnamefont
			{Zheng}}, \bibinfo {author} {\bibfnamefont {N.~S.}\ \bibnamefont {Kiselev}},
		\bibinfo {author} {\bibfnamefont {F.~N.}\ \bibnamefont {Rybakov}}, \bibinfo
		{author} {\bibfnamefont {L.}~\bibnamefont {Yang}}, \bibinfo {author}
		{\bibfnamefont {W.}~\bibnamefont {Shi}}, \bibinfo {author} {\bibfnamefont
			{S.}~\bibnamefont {Bl{\"u}gel}},\ and\ \bibinfo {author} {\bibfnamefont
			{R.~E.}\ \bibnamefont {Dunin-Borkowski}},\ }\href
	{https://doi.org/10.1038/s41586-023-06658-5} {\bibfield  {journal} {\bibinfo
			{journal} {Nature}\ }\textbf {\bibinfo {volume} {623}},\ \bibinfo {pages}
		{718} (\bibinfo {year} {2023})}\BibitemShut {NoStop}%
	\bibitem [{\citenamefont {Duan}\ and\ \citenamefont {Liu}(2004)}]{Duan2004}%
	\BibitemOpen
	\bibfield  {author} {\bibinfo {author} {\bibfnamefont {Y.}~\bibnamefont
			{Duan}}\ and\ \bibinfo {author} {\bibfnamefont {X.}~\bibnamefont {Liu}},\
	}\href {https://doi.org/10.1088/1126-6708/2004/02/028} {\bibfield  {journal}
		{\bibinfo  {journal} {J. High Energy Phys.}\ }\textbf {\bibinfo {volume}
			{2004}}\bibinfo  {number} { (02)},\ \bibinfo {pages} {028}}\BibitemShut
	{NoStop}%
	\bibitem [{\citenamefont {Cirtain}\ \emph {et~al.}(2013)\citenamefont
		{Cirtain}, \citenamefont {Golub}, \citenamefont {Winebarger}, \citenamefont
		{Pontieu}, \citenamefont {Kobayashi}, \citenamefont {Moore}, \citenamefont
		{Walsh}, \citenamefont {Korreck}, \citenamefont {Weber}, \citenamefont
		{McCauley}, \citenamefont {Title}, \citenamefont {Kuzin},\ and\ \citenamefont
		{DeForest}}]{Cirtain2013}%
	\BibitemOpen
	\bibfield  {number} {  }\bibfield  {author} {\bibinfo {author} {\bibfnamefont
			{J.~W.}\ \bibnamefont {Cirtain}}, \bibinfo {author} {\bibfnamefont
			{L.}~\bibnamefont {Golub}}, \bibinfo {author} {\bibfnamefont {A.~R.}\
			\bibnamefont {Winebarger}}, \bibinfo {author} {\bibfnamefont {B.~D.}\
			\bibnamefont {Pontieu}}, \bibinfo {author} {\bibfnamefont {K.}~\bibnamefont
			{Kobayashi}}, \bibinfo {author} {\bibfnamefont {R.~L.}\ \bibnamefont
			{Moore}}, \bibinfo {author} {\bibfnamefont {R.~W.}\ \bibnamefont {Walsh}},
		\bibinfo {author} {\bibfnamefont {K.~E.}\ \bibnamefont {Korreck}}, \bibinfo
		{author} {\bibfnamefont {M.}~\bibnamefont {Weber}}, \bibinfo {author}
		{\bibfnamefont {P.}~\bibnamefont {McCauley}}, \bibinfo {author}
		{\bibfnamefont {A.}~\bibnamefont {Title}}, \bibinfo {author} {\bibfnamefont
			{S.}~\bibnamefont {Kuzin}},\ and\ \bibinfo {author} {\bibfnamefont {C.~E.}\
			\bibnamefont {DeForest}},\ }\href {https://doi.org/10.1038/nature11772}
	{\bibfield  {journal} {\bibinfo  {journal} {Nature}\ }\textbf {\bibinfo
			{volume} {493}},\ \bibinfo {pages} {501} (\bibinfo {year}
		{2013})}\BibitemShut {NoStop}%
	\bibitem [{\citenamefont {Göbel}\ \emph {et~al.}(2021)\citenamefont {Göbel},
		\citenamefont {Mertig},\ and\ \citenamefont {Tretiakov}}]{Gobel2021}%
	\BibitemOpen
	\bibfield  {author} {\bibinfo {author} {\bibfnamefont {B.}~\bibnamefont
			{Göbel}}, \bibinfo {author} {\bibfnamefont {I.}~\bibnamefont {Mertig}},\
		and\ \bibinfo {author} {\bibfnamefont {O.~A.}\ \bibnamefont {Tretiakov}},\
	}\href {https://doi.org/https://doi.org/10.1016/j.physrep.2020.10.001}
	{\bibfield  {journal} {\bibinfo  {journal} {Physics Reports}\ }\textbf
		{\bibinfo {volume} {895}},\ \bibinfo {pages} {1} (\bibinfo {year}
		{2021})}\BibitemShut {NoStop}%
	\bibitem [{\citenamefont {Back}\ \emph {et~al.}(2020)\citenamefont {Back},
		\citenamefont {Cros}, \citenamefont {Ebert}, \citenamefont {Everschor-Sitte},
		\citenamefont {Fert}, \citenamefont {Garst}, \citenamefont {Ma},
		\citenamefont {Mankovsky}, \citenamefont {Monchesky}, \citenamefont
		{Mostovoy}, \citenamefont {Nagaosa}, \citenamefont {Parkin}, \citenamefont
		{Pfleiderer}, \citenamefont {Reyren}, \citenamefont {Rosch}, \citenamefont
		{Taguchi}, \citenamefont {Tokura}, \citenamefont {von Bergmann},\ and\
		\citenamefont {Zang}}]{Back_2020}%
	\BibitemOpen
	\bibfield  {author} {\bibinfo {author} {\bibfnamefont {C.}~\bibnamefont
			{Back}}, \bibinfo {author} {\bibfnamefont {V.}~\bibnamefont {Cros}}, \bibinfo
		{author} {\bibfnamefont {H.}~\bibnamefont {Ebert}}, \bibinfo {author}
		{\bibfnamefont {K.}~\bibnamefont {Everschor-Sitte}}, \bibinfo {author}
		{\bibfnamefont {A.}~\bibnamefont {Fert}}, \bibinfo {author} {\bibfnamefont
			{M.}~\bibnamefont {Garst}}, \bibinfo {author} {\bibfnamefont
			{T.}~\bibnamefont {Ma}}, \bibinfo {author} {\bibfnamefont {S.}~\bibnamefont
			{Mankovsky}}, \bibinfo {author} {\bibfnamefont {T.~L.}\ \bibnamefont
			{Monchesky}}, \bibinfo {author} {\bibfnamefont {M.}~\bibnamefont {Mostovoy}},
		\bibinfo {author} {\bibfnamefont {N.}~\bibnamefont {Nagaosa}}, \bibinfo
		{author} {\bibfnamefont {S.~S.~P.}\ \bibnamefont {Parkin}}, \bibinfo {author}
		{\bibfnamefont {C.}~\bibnamefont {Pfleiderer}}, \bibinfo {author}
		{\bibfnamefont {N.}~\bibnamefont {Reyren}}, \bibinfo {author} {\bibfnamefont
			{A.}~\bibnamefont {Rosch}}, \bibinfo {author} {\bibfnamefont
			{Y.}~\bibnamefont {Taguchi}}, \bibinfo {author} {\bibfnamefont
			{Y.}~\bibnamefont {Tokura}}, \bibinfo {author} {\bibfnamefont
			{K.}~\bibnamefont {von Bergmann}},\ and\ \bibinfo {author} {\bibfnamefont
			{J.}~\bibnamefont {Zang}},\ }\href {https://doi.org/10.1088/1361-6463/ab8418}
	{\bibfield  {journal} {\bibinfo  {journal} {J. Phys. D: Appl. Phys.}\
		}\textbf {\bibinfo {volume} {53}},\ \bibinfo {pages} {363001} (\bibinfo
		{year} {2020})}\BibitemShut {NoStop}%
	\bibitem [{\citenamefont {Fert}\ \emph {et~al.}(2024)\citenamefont {Fert},
		\citenamefont {Ramesh}, \citenamefont {Garcia}, \citenamefont {Casanova},\
		and\ \citenamefont {Bibes}}]{Fert2024}%
	\BibitemOpen
	\bibfield  {author} {\bibinfo {author} {\bibfnamefont {A.}~\bibnamefont
			{Fert}}, \bibinfo {author} {\bibfnamefont {R.}~\bibnamefont {Ramesh}},
		\bibinfo {author} {\bibfnamefont {V.}~\bibnamefont {Garcia}}, \bibinfo
		{author} {\bibfnamefont {F.}~\bibnamefont {Casanova}},\ and\ \bibinfo
		{author} {\bibfnamefont {M.}~\bibnamefont {Bibes}},\ }\href
	{https://doi.org/10.1103/RevModPhys.96.015005} {\bibfield  {journal}
		{\bibinfo  {journal} {Rev. Mod. Phys.}\ }\textbf {\bibinfo {volume} {96}},\
		\bibinfo {pages} {015005} (\bibinfo {year} {2024})}\BibitemShut {NoStop}%
	\bibitem [{\citenamefont {Liu}\ and\ \citenamefont {Nagaosa}(2024)}]{Liu2024}%
	\BibitemOpen
	\bibfield  {author} {\bibinfo {author} {\bibfnamefont {Y.}~\bibnamefont
			{Liu}}\ and\ \bibinfo {author} {\bibfnamefont {N.}~\bibnamefont {Nagaosa}},\
	}\href {https://doi.org/10.1103/PhysRevLett.132.126701} {\bibfield  {journal}
		{\bibinfo  {journal} {Phys. Rev. Lett.}\ }\textbf {\bibinfo {volume} {132}},\
		\bibinfo {pages} {126701} (\bibinfo {year} {2024})}\BibitemShut {NoStop}%
	\bibitem [{\citenamefont {del Ser}\ and\ \citenamefont
		{Lohani}(2023)}]{delSer2023}%
	\BibitemOpen
	\bibfield  {author} {\bibinfo {author} {\bibfnamefont {N.}~\bibnamefont {del
				Ser}}\ and\ \bibinfo {author} {\bibfnamefont {V.}~\bibnamefont {Lohani}},\
	}\href {https://doi.org/10.21468/SciPostPhys.15.2.065} {\bibfield  {journal}
		{\bibinfo  {journal} {SciPost Phys.}\ }\textbf {\bibinfo {volume} {15}},\
		\bibinfo {pages} {065} (\bibinfo {year} {2023})}\BibitemShut {NoStop}%
	\bibitem [{\citenamefont {Tang}\ \emph {et~al.}(2021)\citenamefont {Tang},
		\citenamefont {Wu}, \citenamefont {Wang}, \citenamefont {Kong}, \citenamefont
		{Lv}, \citenamefont {Wei}, \citenamefont {Zang}, \citenamefont {Tian},\ and\
		\citenamefont {Du}}]{Tang2021}%
	\BibitemOpen
	\bibfield  {author} {\bibinfo {author} {\bibfnamefont {J.}~\bibnamefont
			{Tang}}, \bibinfo {author} {\bibfnamefont {Y.}~\bibnamefont {Wu}}, \bibinfo
		{author} {\bibfnamefont {W.}~\bibnamefont {Wang}}, \bibinfo {author}
		{\bibfnamefont {L.}~\bibnamefont {Kong}}, \bibinfo {author} {\bibfnamefont
			{B.}~\bibnamefont {Lv}}, \bibinfo {author} {\bibfnamefont {W.}~\bibnamefont
			{Wei}}, \bibinfo {author} {\bibfnamefont {J.}~\bibnamefont {Zang}}, \bibinfo
		{author} {\bibfnamefont {M.}~\bibnamefont {Tian}},\ and\ \bibinfo {author}
		{\bibfnamefont {H.}~\bibnamefont {Du}},\ }\href
	{https://doi.org/10.1038/s41565-021-00954-9} {\bibfield  {journal} {\bibinfo
			{journal} {Nat. Nanotechnol.}\ }\textbf {\bibinfo {volume} {16}},\ \bibinfo
		{pages} {1086} (\bibinfo {year} {2021})}\BibitemShut {NoStop}%
	\bibitem [{\citenamefont {Zhang}\ \emph {et~al.}(2024)\citenamefont {Zhang},
		\citenamefont {Tang}, \citenamefont {Wu}, \citenamefont {Shi}, \citenamefont
		{Xu}, \citenamefont {Wang}, \citenamefont {Tian},\ and\ \citenamefont
		{Du}}]{Zhang2024}%
	\BibitemOpen
	\bibfield  {author} {\bibinfo {author} {\bibfnamefont {Y.}~\bibnamefont
			{Zhang}}, \bibinfo {author} {\bibfnamefont {J.}~\bibnamefont {Tang}},
		\bibinfo {author} {\bibfnamefont {Y.}~\bibnamefont {Wu}}, \bibinfo {author}
		{\bibfnamefont {M.}~\bibnamefont {Shi}}, \bibinfo {author} {\bibfnamefont
			{X.}~\bibnamefont {Xu}}, \bibinfo {author} {\bibfnamefont {S.}~\bibnamefont
			{Wang}}, \bibinfo {author} {\bibfnamefont {M.}~\bibnamefont {Tian}},\ and\
		\bibinfo {author} {\bibfnamefont {H.}~\bibnamefont {Du}},\ }\href
	{https://doi.org/10.1038/s41467-024-47730-6} {\bibfield  {journal} {\bibinfo
			{journal} {Nat. Commun.}\ }\textbf {\bibinfo {volume} {15}},\ \bibinfo
		{pages} {3391} (\bibinfo {year} {2024})}\BibitemShut {NoStop}%
	\bibitem [{\citenamefont {Barts}\ and\ \citenamefont
		{Mostovoy}(2021)}]{Barts2021}%
	\BibitemOpen
	\bibfield  {author} {\bibinfo {author} {\bibfnamefont {E.}~\bibnamefont
			{Barts}}\ and\ \bibinfo {author} {\bibfnamefont {M.}~\bibnamefont
			{Mostovoy}},\ }\href {https://doi.org/10.1038/s41535-021-00408-4} {\bibfield
		{journal} {\bibinfo  {journal} {npj Quantum Mater.}\ }\textbf {\bibinfo
			{volume} {6}},\ \bibinfo {pages} {104} (\bibinfo {year} {2021})}\BibitemShut
	{NoStop}%
	\bibitem [{\citenamefont {Wang}\ \emph {et~al.}(2019)\citenamefont {Wang},
		\citenamefont {Qaiumzadeh},\ and\ \citenamefont {Brataas}}]{Wang2019}%
	\BibitemOpen
	\bibfield  {author} {\bibinfo {author} {\bibfnamefont {X.~S.}\ \bibnamefont
			{Wang}}, \bibinfo {author} {\bibfnamefont {A.}~\bibnamefont {Qaiumzadeh}},\
		and\ \bibinfo {author} {\bibfnamefont {A.}~\bibnamefont {Brataas}},\ }\href
	{https://doi.org/10.1103/PhysRevLett.123.147203} {\bibfield  {journal}
		{\bibinfo  {journal} {Phys. Rev. Lett.}\ }\textbf {\bibinfo {volume} {123}},\
		\bibinfo {pages} {147203} (\bibinfo {year} {2019})}\BibitemShut {NoStop}%
	\bibitem [{\citenamefont {G\"obel}\ \emph {et~al.}(2020)\citenamefont
		{G\"obel}, \citenamefont {Akosa}, \citenamefont {Tatara},\ and\ \citenamefont
		{Mertig}}]{Gobel2020}%
	\BibitemOpen
	\bibfield  {author} {\bibinfo {author} {\bibfnamefont {B.}~\bibnamefont
			{G\"obel}}, \bibinfo {author} {\bibfnamefont {C.~A.}\ \bibnamefont {Akosa}},
		\bibinfo {author} {\bibfnamefont {G.}~\bibnamefont {Tatara}},\ and\ \bibinfo
		{author} {\bibfnamefont {I.}~\bibnamefont {Mertig}},\ }\href
	{https://doi.org/10.1103/PhysRevResearch.2.013315} {\bibfield  {journal}
		{\bibinfo  {journal} {Phys. Rev. Res.}\ }\textbf {\bibinfo {volume} {2}},\
		\bibinfo {pages} {013315} (\bibinfo {year} {2020})}\BibitemShut {NoStop}%
	\bibitem [{\citenamefont {Liu}\ \emph {et~al.}(2020)\citenamefont {Liu},
		\citenamefont {Hou}, \citenamefont {Han},\ and\ \citenamefont
		{Zang}}]{Liu2020}%
	\BibitemOpen
	\bibfield  {author} {\bibinfo {author} {\bibfnamefont {Y.}~\bibnamefont
			{Liu}}, \bibinfo {author} {\bibfnamefont {W.}~\bibnamefont {Hou}}, \bibinfo
		{author} {\bibfnamefont {X.}~\bibnamefont {Han}},\ and\ \bibinfo {author}
		{\bibfnamefont {J.}~\bibnamefont {Zang}},\ }\href
	{https://doi.org/10.1103/PhysRevLett.124.127204} {\bibfield  {journal}
		{\bibinfo  {journal} {Phys. Rev. Lett.}\ }\textbf {\bibinfo {volume} {124}},\
		\bibinfo {pages} {127204} (\bibinfo {year} {2020})}\BibitemShut {NoStop}%
	\bibitem [{\citenamefont {Raftrey}\ and\ \citenamefont
		{Fischer}(2021)}]{Raftrey2021}%
	\BibitemOpen
	\bibfield  {author} {\bibinfo {author} {\bibfnamefont {D.}~\bibnamefont
			{Raftrey}}\ and\ \bibinfo {author} {\bibfnamefont {P.}~\bibnamefont
			{Fischer}},\ }\href {https://doi.org/10.1103/PhysRevLett.127.257201}
	{\bibfield  {journal} {\bibinfo  {journal} {Phys. Rev. Lett.}\ }\textbf
		{\bibinfo {volume} {127}},\ \bibinfo {pages} {257201} (\bibinfo {year}
		{2021})}\BibitemShut {NoStop}%
	\bibitem [{\citenamefont {Yu}\ \emph {et~al.}(2023)\citenamefont {Yu},
		\citenamefont {Liu}, \citenamefont {Iakoubovskii}, \citenamefont {Nakajima},
		\citenamefont {Kanazawa}, \citenamefont {Nagaosa},\ and\ \citenamefont
		{Tokura}}]{Yu2023}%
	\BibitemOpen
	\bibfield  {author} {\bibinfo {author} {\bibfnamefont {X.}~\bibnamefont
			{Yu}}, \bibinfo {author} {\bibfnamefont {Y.}~\bibnamefont {Liu}}, \bibinfo
		{author} {\bibfnamefont {K.~V.}\ \bibnamefont {Iakoubovskii}}, \bibinfo
		{author} {\bibfnamefont {K.}~\bibnamefont {Nakajima}}, \bibinfo {author}
		{\bibfnamefont {N.}~\bibnamefont {Kanazawa}}, \bibinfo {author}
		{\bibfnamefont {N.}~\bibnamefont {Nagaosa}},\ and\ \bibinfo {author}
		{\bibfnamefont {Y.}~\bibnamefont {Tokura}},\ }\href
	{https://doi.org/https://doi.org/10.1002/adma.202210646} {\bibfield
		{journal} {\bibinfo  {journal} {Adv. Mater.}\ }\textbf {\bibinfo {volume}
			{35}},\ \bibinfo {pages} {2210646} (\bibinfo {year} {2023})}\BibitemShut
	{NoStop}%
	\bibitem [{\citenamefont {Milnor}(1954)}]{Milnor1954}%
	\BibitemOpen
	\bibfield  {author} {\bibinfo {author} {\bibfnamefont {J.}~\bibnamefont
			{Milnor}},\ }\href {http://www.jstor.org/stable/1969685} {\bibfield
		{journal} {\bibinfo  {journal} {Ann. Math.}\ }\textbf {\bibinfo {volume}
			{59}},\ \bibinfo {pages} {177} (\bibinfo {year} {1954})}\BibitemShut
	{NoStop}%
	\bibitem [{\citenamefont {Moffatt}(2013)}]{Moffatt2013}%
	\BibitemOpen
	\bibfield  {author} {\bibinfo {author} {\bibfnamefont {H.~K.}\ \bibnamefont
			{Moffatt}},\ }\href {https://doi.org/10.1073/pnas.1400277111} {\bibfield
		{journal} {\bibinfo  {journal} {Proc. Natl. Acad. Sci.}\ }\textbf {\bibinfo
			{volume} {111}},\ \bibinfo {pages} {3663} (\bibinfo {year}
		{2013})}\BibitemShut {NoStop}%
	\bibitem [{\citenamefont {Machon}(2019)}]{Machon2019}%
	\BibitemOpen
	\bibfield  {author} {\bibinfo {author} {\bibfnamefont {T.}~\bibnamefont
			{Machon}},\ }\href {https://doi.org/10.1080/1358314X.2019.1681113} {\bibfield
		{journal} {\bibinfo  {journal} {Liq. Crys. Today}\ }\textbf {\bibinfo
			{volume} {28}},\ \bibinfo {pages} {58} (\bibinfo {year} {2019})}\BibitemShut
	{NoStop}%
	\bibitem [{\citenamefont {\ifmmode~\check{C}\else \v{C}\fi{}opar}\ and\
		\citenamefont {\ifmmode~\check{Z}\else \v{Z}\fi{}umer}(2011)}]{Copar2011}%
	\BibitemOpen
	\bibfield  {author} {\bibinfo {author} {\bibfnamefont {S.}~\bibnamefont
			{\ifmmode~\check{C}\else \v{C}\fi{}opar}}\ and\ \bibinfo {author}
		{\bibfnamefont {S.}~\bibnamefont {\ifmmode~\check{Z}\else \v{Z}\fi{}umer}},\
	}\href {https://doi.org/10.1103/PhysRevLett.106.177801} {\bibfield  {journal}
		{\bibinfo  {journal} {Phys. Rev. Lett.}\ }\textbf {\bibinfo {volume} {106}},\
		\bibinfo {pages} {177801} (\bibinfo {year} {2011})}\BibitemShut {NoStop}%
	\bibitem [{\citenamefont {Scheeler}\ \emph {et~al.}(2017)\citenamefont
		{Scheeler}, \citenamefont {van Rees}, \citenamefont {Kedia}, \citenamefont
		{Kleckner},\ and\ \citenamefont {Irvine}}]{Scheeler2017}%
	\BibitemOpen
	\bibfield  {author} {\bibinfo {author} {\bibfnamefont {M.~W.}\ \bibnamefont
			{Scheeler}}, \bibinfo {author} {\bibfnamefont {W.~M.}\ \bibnamefont {van
				Rees}}, \bibinfo {author} {\bibfnamefont {H.}~\bibnamefont {Kedia}}, \bibinfo
		{author} {\bibfnamefont {D.}~\bibnamefont {Kleckner}},\ and\ \bibinfo
		{author} {\bibfnamefont {W.~T.~M.}\ \bibnamefont {Irvine}},\ }\href
	{https://doi.org/10.1126/science.aam6897} {\bibfield  {journal} {\bibinfo
			{journal} {Science}\ }\textbf {\bibinfo {volume} {357}},\ \bibinfo {pages}
		{487} (\bibinfo {year} {2017})}\BibitemShut {NoStop}%
	\bibitem [{\citenamefont {Leonov}\ \emph {et~al.}(2016)\citenamefont {Leonov},
		\citenamefont {Monchesky}, \citenamefont {Loudon},\ and\ \citenamefont
		{Bogdanov}}]{Leonov_2016}%
	\BibitemOpen
	\bibfield  {author} {\bibinfo {author} {\bibfnamefont {A.~O.}\ \bibnamefont
			{Leonov}}, \bibinfo {author} {\bibfnamefont {T.~L.}\ \bibnamefont
			{Monchesky}}, \bibinfo {author} {\bibfnamefont {J.~C.}\ \bibnamefont
			{Loudon}},\ and\ \bibinfo {author} {\bibfnamefont {A.~N.}\ \bibnamefont
			{Bogdanov}},\ }\href {https://doi.org/10.1088/0953-8984/28/35/35LT01}
	{\bibfield  {journal} {\bibinfo  {journal} {J. Phys. Condens. Matter}\
		}\textbf {\bibinfo {volume} {28}},\ \bibinfo {pages} {35LT01} (\bibinfo
		{year} {2016})}\BibitemShut {NoStop}%
	\bibitem [{\citenamefont {Rybakov}\ \emph {et~al.}(2015)\citenamefont
		{Rybakov}, \citenamefont {Borisov}, \citenamefont {Bl\"ugel},\ and\
		\citenamefont {Kiselev}}]{Rybakov2015}%
	\BibitemOpen
	\bibfield  {author} {\bibinfo {author} {\bibfnamefont {F.~N.}\ \bibnamefont
			{Rybakov}}, \bibinfo {author} {\bibfnamefont {A.~B.}\ \bibnamefont
			{Borisov}}, \bibinfo {author} {\bibfnamefont {S.}~\bibnamefont {Bl\"ugel}},\
		and\ \bibinfo {author} {\bibfnamefont {N.~S.}\ \bibnamefont {Kiselev}},\
	}\href {https://doi.org/10.1103/PhysRevLett.115.117201} {\bibfield  {journal}
		{\bibinfo  {journal} {Phys. Rev. Lett.}\ }\textbf {\bibinfo {volume} {115}},\
		\bibinfo {pages} {117201} (\bibinfo {year} {2015})}\BibitemShut {NoStop}%
	\bibitem [{\citenamefont {Kuchkin}\ \emph {et~al.}(2023)\citenamefont
		{Kuchkin}, \citenamefont {Kiselev}, \citenamefont {Rybakov}, \citenamefont
		{Lobanov}, \citenamefont {Blügel},\ and\ \citenamefont
		{Uzdin}}]{Kuchkin2023}%
	\BibitemOpen
	\bibfield  {author} {\bibinfo {author} {\bibfnamefont {V.~M.}\ \bibnamefont
			{Kuchkin}}, \bibinfo {author} {\bibfnamefont {N.~S.}\ \bibnamefont
			{Kiselev}}, \bibinfo {author} {\bibfnamefont {F.~N.}\ \bibnamefont
			{Rybakov}}, \bibinfo {author} {\bibfnamefont {I.~S.}\ \bibnamefont
			{Lobanov}}, \bibinfo {author} {\bibfnamefont {S.}~\bibnamefont {Blügel}},\
		and\ \bibinfo {author} {\bibfnamefont {V.~M.}\ \bibnamefont {Uzdin}},\ }\href
	{https://www.frontiersin.org/journals/physics/articles/10.3389/fphy.2023.1201018}
	{\bibfield  {journal} {\bibinfo  {journal} {Frontiers in Physics}\ }\textbf
		{\bibinfo {volume} {11}} (\bibinfo {year} {2023})}\BibitemShut {NoStop}%
	\bibitem [{\citenamefont {Ackerman}\ and\ \citenamefont
		{Smalyukh}(2017)}]{Ackerman17}%
	\BibitemOpen
	\bibfield  {author} {\bibinfo {author} {\bibfnamefont {P.~J.}\ \bibnamefont
			{Ackerman}}\ and\ \bibinfo {author} {\bibfnamefont {I.~I.}\ \bibnamefont
			{Smalyukh}},\ }\href {https://doi.org/10.1103/physrevx.7.011006} {\bibfield
		{journal} {\bibinfo  {journal} {Phys. Rev. X}\ }\textbf {\bibinfo {volume}
			{7}},\ \bibinfo {pages} {011006} (\bibinfo {year} {2017})}\BibitemShut
	{NoStop}%
	\bibitem [{\citenamefont {Samoilenka}\ and\ \citenamefont
		{Shnir}(2017)}]{Samoilenka2017}%
	\BibitemOpen
	\bibfield  {author} {\bibinfo {author} {\bibfnamefont {A.}~\bibnamefont
			{Samoilenka}}\ and\ \bibinfo {author} {\bibfnamefont {Y.}~\bibnamefont
			{Shnir}},\ }\href {https://doi.org/10.1007/JHEP09(2017)029} {\bibfield
		{journal} {\bibinfo  {journal} {J. High Energy Phys.}\ }\textbf {\bibinfo
			{volume} {2017}}\bibinfo  {number} { (9)},\ \bibinfo {pages}
		{29}}\BibitemShut {NoStop}%
	\bibitem [{\citenamefont {Tai}\ \emph {et~al.}(2018)\citenamefont {Tai},
		\citenamefont {Ackerman},\ and\ \citenamefont {Smalyukh}}]{Tai2018}%
	\BibitemOpen
	\bibfield  {number} {  }\bibfield  {author} {\bibinfo {author} {\bibfnamefont
			{J.-S.~B.}\ \bibnamefont {Tai}}, \bibinfo {author} {\bibfnamefont {P.~J.}\
			\bibnamefont {Ackerman}},\ and\ \bibinfo {author} {\bibfnamefont {I.~I.}\
			\bibnamefont {Smalyukh}},\ }\href {https://doi.org/10.1073/pnas.1716887115}
	{\bibfield  {journal} {\bibinfo  {journal} {Proc. Natl. Acad. Sci.}\ }\textbf
		{\bibinfo {volume} {115}},\ \bibinfo {pages} {921} (\bibinfo {year}
		{2018})}\BibitemShut {NoStop}%
	\bibitem [{\citenamefont {Auckly}\ and\ \citenamefont
		{Kapitanski}(2005)}]{Auckly2005}%
	\BibitemOpen
	\bibfield  {author} {\bibinfo {author} {\bibfnamefont {D.}~\bibnamefont
			{Auckly}}\ and\ \bibinfo {author} {\bibfnamefont {L.}~\bibnamefont
			{Kapitanski}},\ }\href {https://doi.org/10.1007/s00220-005-1289-6} {\bibfield
		{journal} {\bibinfo  {journal} {Commun. Math. Phys.}\ }\textbf {\bibinfo
			{volume} {256}},\ \bibinfo {pages} {611} (\bibinfo {year}
		{2005})}\BibitemShut {NoStop}%
	\bibitem [{\citenamefont {J\"aykk\"a}\ and\ \citenamefont
		{Hietarinta}(2009)}]{Hietarinta2009}%
	\BibitemOpen
	\bibfield  {author} {\bibinfo {author} {\bibfnamefont {J.}~\bibnamefont
			{J\"aykk\"a}}\ and\ \bibinfo {author} {\bibfnamefont {J.}~\bibnamefont
			{Hietarinta}},\ }\href {https://doi.org/10.1103/PhysRevD.79.125027}
	{\bibfield  {journal} {\bibinfo  {journal} {Phys. Rev. D}\ }\textbf {\bibinfo
			{volume} {79}},\ \bibinfo {pages} {125027} (\bibinfo {year}
		{2009})}\BibitemShut {NoStop}%
	\bibitem [{\citenamefont {Kobayashi}\ and\ \citenamefont
		{Nitta}(2013)}]{Kobayashi2013}%
	\BibitemOpen
	\bibfield  {author} {\bibinfo {author} {\bibfnamefont {M.}~\bibnamefont
			{Kobayashi}}\ and\ \bibinfo {author} {\bibfnamefont {M.}~\bibnamefont
			{Nitta}},\ }\href
	{https://doi.org/https://doi.org/10.1016/j.nuclphysb.2013.08.012} {\bibfield
		{journal} {\bibinfo  {journal} {Nucl. Phys. B}\ }\textbf {\bibinfo {volume}
			{876}},\ \bibinfo {pages} {605} (\bibinfo {year} {2013})}\BibitemShut
	{NoStop}%
	\bibitem [{\citenamefont {Balakrishnan}\ \emph {et~al.}(2023)\citenamefont
		{Balakrishnan}, \citenamefont {Dandoloff},\ and\ \citenamefont
		{Saxena}}]{BALAKRISHNAN2023}%
	\BibitemOpen
	\bibfield  {author} {\bibinfo {author} {\bibfnamefont {R.}~\bibnamefont
			{Balakrishnan}}, \bibinfo {author} {\bibfnamefont {R.}~\bibnamefont
			{Dandoloff}},\ and\ \bibinfo {author} {\bibfnamefont {A.}~\bibnamefont
			{Saxena}},\ }\href
	{https://doi.org/https://doi.org/10.1016/j.physleta.2023.128975} {\bibfield
		{journal} {\bibinfo  {journal} {Phys. Lett. A}\ }\textbf {\bibinfo {volume}
			{480}},\ \bibinfo {pages} {128975} (\bibinfo {year} {2023})}\BibitemShut
	{NoStop}%
	\bibitem [{\citenamefont {Knapman}\ \emph
		{et~al.}(2024{\natexlab{a}})\citenamefont {Knapman}, \citenamefont
		{Tausendpfund}, \citenamefont {D\'iaz},\ and\ \citenamefont
		{Everschor-Sitte}}]{Knapman2024}%
	\BibitemOpen
	\bibfield  {author} {\bibinfo {author} {\bibfnamefont {R.}~\bibnamefont
			{Knapman}}, \bibinfo {author} {\bibfnamefont {T.}~\bibnamefont
			{Tausendpfund}}, \bibinfo {author} {\bibfnamefont {S.~A.}\ \bibnamefont
			{D\'iaz}},\ and\ \bibinfo {author} {\bibfnamefont {K.}~\bibnamefont
			{Everschor-Sitte}},\ }\href {https://doi.org/10.1038/s42005-024-01628-3}
	{\bibfield  {journal} {\bibinfo  {journal} {Commun. Phys.}\ }\textbf
		{\bibinfo {volume} {7}},\ \bibinfo {pages} {151} (\bibinfo {year}
		{2024}{\natexlab{a}})}\BibitemShut {NoStop}%
	\bibitem [{\citenamefont {Berger}\ and\ \citenamefont
		{Field}(1984)}]{Berger1984}%
	\BibitemOpen
	\bibfield  {author} {\bibinfo {author} {\bibfnamefont {M.~A.}\ \bibnamefont
			{Berger}}\ and\ \bibinfo {author} {\bibfnamefont {G.~B.}\ \bibnamefont
			{Field}},\ }\href {https://doi.org/10.1017/S0022112084002019} {\bibfield
		{journal} {\bibinfo  {journal} {J. Fluid Mech.}\ }\textbf {\bibinfo {volume}
			{147}},\ \bibinfo {pages} {133–148} (\bibinfo {year} {1984})}\BibitemShut
	{NoStop}%
	\bibitem [{\citenamefont {Prior}\ and\ \citenamefont
		{Yeates}(2014)}]{Prior2014}%
	\BibitemOpen
	\bibfield  {author} {\bibinfo {author} {\bibfnamefont {C.}~\bibnamefont
			{Prior}}\ and\ \bibinfo {author} {\bibfnamefont {A.~R.}\ \bibnamefont
			{Yeates}},\ }\href {https://doi.org/10.1088/0004-637X/787/2/100} {\bibfield
		{journal} {\bibinfo  {journal} {The Astrophysical Journal}\ }\textbf
		{\bibinfo {volume} {787}},\ \bibinfo {pages} {100} (\bibinfo {year}
		{2014})}\BibitemShut {NoStop}%
	\bibitem [{\citenamefont {Moreau}(1961)}]{Moreau1961}%
	\BibitemOpen
	\bibfield  {author} {\bibinfo {author} {\bibfnamefont {J.~J.}\ \bibnamefont
			{Moreau}},\ }\href {https://hal.science/hal-01865239} {\bibfield  {journal}
		{\bibinfo  {journal} {{Comptes rendus hebdomadaires des s{\'e}ances de
					l'Acad{\'e}mie des sciences}}\ }\textbf {\bibinfo {volume} {252}},\ \bibinfo
		{pages} {2810} (\bibinfo {year} {1961})}\BibitemShut {NoStop}%
	\bibitem [{\citenamefont {Moffatt}(1969)}]{Moffatt1969}%
	\BibitemOpen
	\bibfield  {author} {\bibinfo {author} {\bibfnamefont {H.~K.}\ \bibnamefont
			{Moffatt}},\ }\href {https://doi.org/10.1017/S0022112069000991} {\bibfield
		{journal} {\bibinfo  {journal} {Journal of Fluid Mechanics}\ }\textbf
		{\bibinfo {volume} {35}},\ \bibinfo {pages} {117–129} (\bibinfo {year}
		{1969})}\BibitemShut {NoStop}%
	\bibitem [{\citenamefont {Moffatt}(1978)}]{moffatt1978}%
	\BibitemOpen
	\bibfield  {author} {\bibinfo {author} {\bibfnamefont {H.}~\bibnamefont
			{Moffatt}},\ }\href
	{https://books.google.de/books/about/Magnetic_Field_Generation_in_Electricall.html?id=cAo4AAAAIAAJ}
	{\emph {\bibinfo {title} {Magnetic Field Generation in Electrically
				Conducting Fluids}}}\ (\bibinfo  {publisher} {Cambridge University Press},\
	\bibinfo {year} {1978})\BibitemShut {NoStop}%
	\bibitem [{\citenamefont {Moffatt}(1981)}]{Moffatt1981}%
	\BibitemOpen
	\bibfield  {author} {\bibinfo {author} {\bibfnamefont {H.~K.}\ \bibnamefont
			{Moffatt}},\ }\href {https://doi.org/10.1017/S002211208100150X} {\bibfield
		{journal} {\bibinfo  {journal} {Journal of Fluid Mechanics}\ }\textbf
		{\bibinfo {volume} {106}},\ \bibinfo {pages} {27–47} (\bibinfo {year}
		{1981})}\BibitemShut {NoStop}%
	\bibitem [{Note1()}]{Note1}%
	\BibitemOpen
	\bibinfo {note} {Please note that every solenoidal vector field can be
		decomposed into flux tubes. This decomposition is not necessarily unique, but
		all decompositions lead to the same result in Eq.~\protect \eqref
		{eq:Hopf_decomposed}. For a definition of flux tubes we refer to Ref.~\cite
		{Prior2014}.}\BibitemShut {Stop}%
	\bibitem [{\citenamefont {Moffatt}\ and\ \citenamefont
		{Ricca}(1992)}]{Moffat1992}%
	\BibitemOpen
	\bibfield  {author} {\bibinfo {author} {\bibfnamefont {H.~K.}\ \bibnamefont
			{Moffatt}}\ and\ \bibinfo {author} {\bibfnamefont {R.~L.}\ \bibnamefont
			{Ricca}},\ }\href {https://doi.org/10.1098/rspa.1992.0159} {\bibfield
		{journal} {\bibinfo  {journal} {Proc. Roy. Soc. Lond. A}\ }\textbf {\bibinfo
			{volume} {439}},\ \bibinfo {pages} {411} (\bibinfo {year}
		{1992})}\BibitemShut {NoStop}%
	\bibitem [{\citenamefont {Ricca}\ and\ \citenamefont
		{Moffatt}(1992)}]{Ricca1992}%
	\BibitemOpen
	\bibfield  {author} {\bibinfo {author} {\bibfnamefont {R.~L.}\ \bibnamefont
			{Ricca}}\ and\ \bibinfo {author} {\bibfnamefont {H.~K.}\ \bibnamefont
			{Moffatt}},\ }\bibinfo {title} {The helicity of a knotted vortex filament},\
	in\ \href {https://doi.org/10.1007/978-94-017-3550-6_11} {\emph {\bibinfo
			{booktitle} {Topological Aspects of the Dynamics of Fluids and Plasmas}}},\
	\bibinfo {editor} {edited by\ \bibinfo {editor} {\bibfnamefont {H.~K.}\
			\bibnamefont {Moffatt}}, \bibinfo {editor} {\bibfnamefont {G.~M.}\
			\bibnamefont {Zaslavsky}}, \bibinfo {editor} {\bibfnamefont {P.}~\bibnamefont
			{Comte}},\ and\ \bibinfo {editor} {\bibfnamefont {M.}~\bibnamefont {Tabor}}}\
	(\bibinfo  {publisher} {Springer Netherlands},\ \bibinfo {address}
	{Dordrecht},\ \bibinfo {year} {1992})\ pp.\ \bibinfo {pages}
	{225--236}\BibitemShut {NoStop}%
	\bibitem [{Note2()}]{Note2}%
	\BibitemOpen
	\bibinfo {note} {Equation \protect \eqref {eq:linkingnum} contains integrals
		over closed curves. For field lines that run vertically along the
		$z$-direction, the periodic boundary conditions imply that the field lines
		close trivially at infinity. An alternative is to define ``winding numbers''
		of open field lines as in Ref.~\cite {Prior2014}}\BibitemShut {NoStop}%
	\bibitem [{Note3()}]{Note3}%
	\BibitemOpen
	\bibinfo {note} {Note that, in general, linking numbers can be written as the
		sum of twist and writhe components~\cite
		{Moffatt2013,Machon2019,Copar2011,Scheeler2017}.}\BibitemShut {Stop}%
	\bibitem [{\citenamefont {Whitehead}(1947)}]{whitehead1947}%
	\BibitemOpen
	\bibfield  {author} {\bibinfo {author} {\bibfnamefont {J.}~\bibnamefont
			{Whitehead}},\ }\href {https://doi.org/10.1073/pnas.33.5.117} {\bibfield
		{journal} {\bibinfo  {journal} {PNAS}\ }\textbf {\bibinfo {volume} {33}},\
		\bibinfo {pages} {117} (\bibinfo {year} {1947})}\BibitemShut {NoStop}%
	\bibitem [{\citenamefont {Knapman}\ \emph
		{et~al.}(2024{\natexlab{b}})\citenamefont {Knapman}, \citenamefont {Azhar},
		\citenamefont {Pignedoli}, \citenamefont {Gallard}, \citenamefont {Hertel},
		\citenamefont {Leliaert},\ and\ \citenamefont
		{Everschor-Sitte}}]{Knapman2024a}%
	\BibitemOpen
	\bibfield  {author} {\bibinfo {author} {\bibfnamefont {R.}~\bibnamefont
			{Knapman}}, \bibinfo {author} {\bibfnamefont {M.}~\bibnamefont {Azhar}},
		\bibinfo {author} {\bibfnamefont {A.}~\bibnamefont {Pignedoli}}, \bibinfo
		{author} {\bibfnamefont {L.}~\bibnamefont {Gallard}}, \bibinfo {author}
		{\bibfnamefont {R.}~\bibnamefont {Hertel}}, \bibinfo {author} {\bibfnamefont
			{J.}~\bibnamefont {Leliaert}},\ and\ \bibinfo {author} {\bibfnamefont
			{K.}~\bibnamefont {Everschor-Sitte}},\ }\href
	{https://arxiv.org/abs/2410.22058} {\bibfield  {journal} {\bibinfo  {journal}
			{arXiv preprint arXiv:2410.22058}\ } (\bibinfo {year}
		{2024}{\natexlab{b}})}\BibitemShut {NoStop}%
	\bibitem [{\citenamefont {Volovik}(1987)}]{Volovik1987}%
	\BibitemOpen
	\bibfield  {author} {\bibinfo {author} {\bibfnamefont {G.~E.}\ \bibnamefont
			{Volovik}},\ }\href {https://doi.org/10.1088/0022-3719/20/7/003} {\bibfield
		{journal} {\bibinfo  {journal} {Journal of Physics C: Solid State Physics}\
		}\textbf {\bibinfo {volume} {20}},\ \bibinfo {pages} {L83} (\bibinfo {year}
		{1987})}\BibitemShut {NoStop}%
	\bibitem [{\citenamefont {Everschor-Sitte}\ and\ \citenamefont
		{Sitte}(2014)}]{Everschor2014}%
	\BibitemOpen
	\bibfield  {author} {\bibinfo {author} {\bibfnamefont {K.}~\bibnamefont
			{Everschor-Sitte}}\ and\ \bibinfo {author} {\bibfnamefont {M.}~\bibnamefont
			{Sitte}},\ }\href {https://doi.org/10.1063/1.4870695} {\bibfield  {journal}
		{\bibinfo  {journal} {J. Appl. Phys.}\ }\textbf {\bibinfo {volume} {115}},\
		\bibinfo {pages} {172602} (\bibinfo {year} {2014})}\BibitemShut {NoStop}%
	\bibitem [{\citenamefont {Barnett}\ \emph {et~al.}(2023)\citenamefont
		{Barnett}, \citenamefont {Speirits},\ and\ \citenamefont
		{Götte}}]{Barnett_2023}%
	\BibitemOpen
	\bibfield  {author} {\bibinfo {author} {\bibfnamefont {S.~M.}\ \bibnamefont
			{Barnett}}, \bibinfo {author} {\bibfnamefont {F.~C.}\ \bibnamefont
			{Speirits}},\ and\ \bibinfo {author} {\bibfnamefont {J.~B.}\ \bibnamefont
			{Götte}},\ }\href {https://doi.org/10.1209/0295-5075/ace8b7} {\bibfield
		{journal} {\bibinfo  {journal} {Europhys. Lett.}\ }\textbf {\bibinfo {volume}
			{143}},\ \bibinfo {pages} {35002} (\bibinfo {year} {2023})}\BibitemShut
	{NoStop}%
	\bibitem [{Note4()}]{Note4}%
	\BibitemOpen
	\bibinfo {note} {(referred to as sd$_{1}^{+}$ and sd$_{-1}^{+}$ in Ref.~\cite
		{Azhar2022})}\BibitemShut {NoStop}%
	\bibitem [{Note5()}]{Note5}%
	\BibitemOpen
	\bibinfo {note} {(referred to as sd$_{1}^{\protect \mathrm {Sk-}}$ in
		Ref.~\cite {Azhar2022})}\BibitemShut {NoStop}%
	\bibitem [{\citenamefont {del Ser}\ \emph {et~al.}(2024)\citenamefont {del
			Ser}, \citenamefont {El~Achchi},\ and\ \citenamefont {Rosch}}]{delSer2024}%
	\BibitemOpen
	\bibfield  {author} {\bibinfo {author} {\bibfnamefont {N.}~\bibnamefont {del
				Ser}}, \bibinfo {author} {\bibfnamefont {I.}~\bibnamefont {El~Achchi}},\ and\
		\bibinfo {author} {\bibfnamefont {A.}~\bibnamefont {Rosch}},\ }\href
	{https://doi.org/10.1103/PhysRevB.110.094442} {\bibfield  {journal} {\bibinfo
			{journal} {Phys. Rev. B}\ }\textbf {\bibinfo {volume} {110}},\ \bibinfo
		{pages} {094442} (\bibinfo {year} {2024})}\BibitemShut {NoStop}%
	\bibitem [{\citenamefont {Dzyaloshinkskii}(1964)}]{Dzyaloshinkskii1964}%
	\BibitemOpen
	\bibfield  {author} {\bibinfo {author} {\bibfnamefont {I.~E.}\ \bibnamefont
			{Dzyaloshinkskii}},\ }\href@noop {} {\bibfield  {journal} {\bibinfo
			{journal} {J. Exptl. Theoret. Phys. (U.S.S.R.)}\ }\textbf {\bibinfo {volume}
			{46}},\ \bibinfo {pages} {1420} (\bibinfo {year} {1964})}\BibitemShut
	{NoStop}%
	\bibitem [{\citenamefont {Vansteenkiste}\ \emph {et~al.}(2014)\citenamefont
		{Vansteenkiste}, \citenamefont {Leliaert}, \citenamefont {Dvornik},
		\citenamefont {Helsen}, \citenamefont {Garcia-Sanchez},\ and\ \citenamefont
		{{Van Waeyenberge}}}]{Vansteenkiste2014}%
	\BibitemOpen
	\bibfield  {author} {\bibinfo {author} {\bibfnamefont {A.}~\bibnamefont
			{Vansteenkiste}}, \bibinfo {author} {\bibfnamefont {J.}~\bibnamefont
			{Leliaert}}, \bibinfo {author} {\bibfnamefont {M.}~\bibnamefont {Dvornik}},
		\bibinfo {author} {\bibfnamefont {M.}~\bibnamefont {Helsen}}, \bibinfo
		{author} {\bibfnamefont {F.}~\bibnamefont {Garcia-Sanchez}},\ and\ \bibinfo
		{author} {\bibfnamefont {B.}~\bibnamefont {{Van Waeyenberge}}},\ }\href
	{https://doi.org/10.1063/1.4899186} {\bibfield  {journal} {\bibinfo
			{journal} {AIP Adv.}\ }\textbf {\bibinfo {volume} {4}},\ \bibinfo {pages}
		{107133} (\bibinfo {year} {2014})}\BibitemShut {NoStop}%
	\bibitem [{\citenamefont {Kravchuk}(2023)}]{Kravchuk2023}%
	\BibitemOpen
	\bibfield  {author} {\bibinfo {author} {\bibfnamefont {V.}~\bibnamefont
			{Kravchuk}},\ }\href {https://doi.org/10.13140/RG.2.2.22653.74728} {\bibinfo
		{title} {Problems in micromagnetism}} (\bibinfo {year} {2023})\BibitemShut
	{NoStop}%
\end{thebibliography}
\end{document}